\documentclass[conference]{IEEEtran}
\usepackage{cite}
\usepackage{amsmath,amssymb,amsfonts}
\usepackage{algorithmic}
\usepackage{graphicx}
\usepackage{textcomp}
\usepackage{xcolor}
\usepackage[hyphens]{url}

\def\BibTeX{{\rm B\kern-.05em{\sc i\kern-.025em b}\kern-.08em
    T\kern-.1667em\lower.7ex\hbox{E}\kern-.125emX}}

\usepackage{xspace}
\usepackage{kotex}
\usepackage{lipsum}
\usepackage{enumitem}
\usepackage{fancyhdr}
\usepackage{hyperref}
\usepackage{comment}
\usepackage{makecell}
\usepackage{tabularx}
\usepackage{multirow}
\usepackage{comment}
\usepackage{xfrac}
\usepackage{multirow}
\usepackage{tcolorbox}
\usepackage{booktabs}
\usepackage{color,soul}
\usepackage[utf8]{inputenc}
\usepackage[english]{babel}
\usepackage{balance}
\usepackage{circledsteps}

\newcommand{\rdgranul}{$\text{RD}_\text{data}$\xspace}
\newcommand{\hcnt}{$\text{H}_\text{cnt}$\xspace}
\newcommand{\name}{DRAMScope\xspace}
\newcommand{\sixf}{$\text{6F}^\text{2}$\xspace}

\newcommand{\aib}{AIB\xspace}
\newcommand{\aibs}{AIBs\xspace}

\newcommand{\eg}{\emph{e.g.}\xspace}
\newcommand{\ie}{\emph{i.e.}\xspace}
\def\cf{\textit{cf}.\xspace}

\newcommand{\samsung}{Mfr. A\xspace}
\newcommand{\hynix}{Mfr. B\xspace}
\newcommand{\micron}{Mfr. C\xspace}

\newcommand{\jaemin}[1]{{\color{blue}#1}}
\newcommand{\minbok}[1]{{\color{orange}#1}}

\newcommand{\rev}[1]{{\color{blue}#1}}

\title{DRAMScope: Uncovering DRAM Microarchitecture and Characteristics by Issuing Memory Commands}

\author{\IEEEauthorblockN{Hwayong Nam\IEEEauthorrefmark{2},
Seungmin Baek\IEEEauthorrefmark{2},
Minbok Wi\IEEEauthorrefmark{2},
Michael Jaemin Kim\IEEEauthorrefmark{2},
Jaehyun Park\IEEEauthorrefmark{2},
Chihun Song\IEEEauthorrefmark{3},\\
Nam Sung Kim\IEEEauthorrefmark{3},
Jung Ho Ahn\IEEEauthorrefmark{2}}
\IEEEauthorblockA{\IEEEauthorrefmark{2}Seoul National University, \IEEEauthorrefmark{3}University of Illinois at Urbana-Champaign}
\IEEEauthorblockN{
\IEEEauthorrefmark{2}\{nhy4916, qortmdalss, homakaka, michael604, wogus20002, gajh\}@snu.ac.kr, 
\IEEEauthorrefmark{3}\{chihuns2, nskim\}@illinois.edu
}
}

\twocolumn

\begin{document}
\maketitle

\begin{abstract}
The demand for precise information on DRAM microarchitectures and error characteristics has surged, driven by the need to explore processing in memory, enhance reliability, and mitigate security vulnerability. 
Nonetheless, DRAM manufacturers have disclosed only a limited amount of information, making it difficult to find specific information on their DRAM microarchitectures. 
This paper addresses this gap by presenting more rigorous findings on the microarchitectures of commodity DRAM chips and their impacts on the characteristics of activate-induced bitflips (AIBs), such as RowHammer and RowPress.
The previous studies have also attempted to understand the DRAM microarchitectures and associated behaviors, but we have found some of their results to be misled by inaccurate address mapping and internal data swizzling, or lack of a deeper understanding of the modern DRAM cell structure.
%
For accurate and efficient reverse-engineering, we use three tools: AIBs, retention time test, and RowCopy, which can be cross-validated. 
With these three tools, we first take a macroscopic view of modern DRAM chips to uncover the size, structure, and operation of their subarrays, memory array tiles (MATs), and rows.
Then, we analyze AIB characteristics based on the microscopic view of the DRAM microarchitecture, such as $\text{6F}^\text{2}$ cell layout, through which we rectify misunderstandings regarding AIBs and discover a new data pattern that accelerates AIBs. 
%
Lastly, based on our findings at both macroscopic and microscopic levels, we identify previously unknown AIB vulnerabilities and propose a simple yet effective protection solution. 
\end{abstract}

\section{Introduction}
\label{sec:1_introduction}

A deep understanding of DRAM microarchitecture and error characteristics is more important than ever; processing in memory (PIM) spotlighted
~\cite{isscc-2022-aim, isca-2021-hbm-pim, micro-2021-trim, micro-2017-ambit, micro-2019-computedram},
soft/hard error rate exacerbated~\cite{asplos-2015-dramerrors,hpca-2023-dramfault, micro-2023-dram-fault}, and yet another activate-induced bitflip (AIB) vulnerability discovered~\cite{isca-2023-rowpress}.
For instance, constructing secure and efficient \aib protection solutions without an accurate understanding of DRAM error behaviors linked to specific aspects of a DRAM microarchitecture would be undoubtedly challenging. 
Likewise, a detailed knowledge of the DRAM microarchitecture is essential in exploring efficient PIM architectures.
However, the DRAM microarchitecture has undergone decades of optimizations to improve not only the cell density or energy efficiency but also the manufacturing yield and cost.
Such optimizations are manufacturer-specific and proprietary~\cite{irps-2022-secrecy}, significantly hindering efforts to uncover the true DRAM microarchitecture and error characteristics.

To fill this critical gap, a large body of prior work has exploited creative reverse-engineering methodologies. 
They have relied on scarcely disclosed knowledge or assumptions~\cite{jedec-ddr4,jedec-hbm, msr-2021-mfit,jedec-ddr4-rcd,micron-dimm16GB,micron-dimm32GB} to uncover error characteristics~\cite{isca-2013-retention,isca-2014-flipping,isca-2020-revisit,isca-2023-rowpress,micro-2021-deeper,msr-2021-mfit}, undefined DRAM operations~\cite{micro-2013-rowclone,micro-2022-hira}, microarchitectural components transparent to memory controllers, such as \aib protection solutions~\cite{sp-2020-trrespass,micro-2021-uncovering} or on-die ECC~\cite{micro-2020-beer,micro-2021-harp}, to list a few.
Nonetheless, we have found a number of previous efforts to discover the DRAM microarchitecture are limited in scope, outdated, or even misleading due to an insufficient understanding of the modern DRAM \sixf cell structure (see Figure~\ref{fig:microscopic}), complex mapping of CPU physical addresses to DRAM addresses, and swizzling of CPU data within DRAM.\footnote{DRAM internal data swizzling occurs as data collected from the subarray is reorganized to get transferred to the CPU. See \S~\ref{sec:4_contribution1}.}

In this paper, we conduct a comprehensive study to better understand the DRAM microarchitecture (\emph{macroscopic} level) and \aib characteristics (\emph{microscopic} level) of modern DRAM chips, leveraging three different reverse-engineering techniques and our recent knowledge of the aforementioned address mapping and data swizzling.
%
Without a thorough understanding of the address mapping and data swizzling, attempting to control DRAM chips can lead to inconsistencies between the user's intended access and the physical access.
Similarly, comprehending the \sixf cell structure and the physical distances between cells and intervening gate types is essential for obtaining clearer insights from reverse-engineering efforts.
We uniquely exploit this interplay by utilizing DRAM errors to uncover the DRAM microarchitecture while simultaneously leveraging our recent microarchitectural knowledge to investigate error characteristics.
%
%
%

\noindent
\textbf{Reliable and cross-validatable reverse-engineering techniques (\S\ref{sec:3_experiments}):}
To reverse-engineer the DRAM microarchitecture without intrusive measures such as physical probing~\cite{sp-2019-eccploit,msr-2021-mfit}, we use three techniques using standard DRAM commands in a controlled FPGA-based environment. The three techniques are as follows: (1) causing \aibs such as RowHammer~\cite{isca-2014-flipping} and RowPress~\cite{isca-2023-rowpress}, (2) performing in-memory row copy operations (RowCopy)~\cite{micro-2013-rowclone,micro-2019-computedram}, and (3) inducing data retention errors.
%
Analyzing the results obtained from these three techniques provides us with not only the accurate error characteristics but also the hidden details of the DRAM microarchitecture.
Furthermore, we highlight the challenges posed by intricate address mapping and data swizzling schemes, including row address remapping at individual DRAM chips, row address inversion at the registered clock driver (RCD) chip, and data pin (DQ) twisting.
Although such information is often publicly disclosed~\cite{jedec-ddr4-rcd,jedec-ddr4-rdimm}, they are scattered across documents and can easily be omitted, leading to incorrect analysis.

\noindent
\textbf{Macroscopic DRAM microarchitectural analysis (\S\ref{sec:4_contribution1}):}
We conduct a \emph{macroscopic} analysis that does not require knowledge of the \sixf cell structure to reverse-engineer the \emph{data swizzling} and identify previously unreported structural \textbf{\underline{O}}bservations at the subarray, row, and memory array tile (MAT) levels.
\textbf{(O1)} We reconstruct the DRAM chip internal data swizzling based on our observation that horizontally adjacent victim cells affect \aib, which we elaborate later in \S\ref{sec:5_contribution2}.
We observe that data within a single read are reorganized and collected from multiple MATs.
\textbf{(O2)} We also identify the MAT \emph{width}, or the number of cells constituting a single row in a MAT.
\textbf{(O3)} For certain DRAM chips, two separate rows specified in physical address are coupled and activated together by a single row command.
\textbf{(O4)} For all tested DRAM modules, the number of rows in a single subarray (\emph{height}) is not a power of two, and multiple subarray heights can coexist even in a single chip.
Also, we recognize a clear trend of increase in subarray height over DRAM generations.
\textbf{(O5)} Certain DRAM chips combine two subarrays at the physical edge to work in tandem (edge subarrays), which is deducible from the open bitline structure~\cite{2010-DRAM_book}. 
\textbf{(O6)} The bit error rate (BER) by \aib is lower in edge subarrays, possibly due to the dummy bitlines.\footnote{The edge subarrays of an open bitline structure utilize only half the bitlines, leaving the other half as dummy bitlines.} 

%

\noindent
\textbf{Microscopic DRAM error analysis (\S\ref{sec:5_contribution2}):}
With our \emph{microscopic} analysis that exploits our knowledge of the \sixf cell structure, we present the following observations. 
\textbf{(O7)} There exists an alternating pattern in RowPress vulnerability, which reverses when either the row parity (even/odd) changes or the aggressor direction (up/down) changes.
This faithfully reflects the \sixf cell structure.
\textbf{(O8)} RowHammer also exhibits a similar alternating pattern, which is reversed when row parity, aggressor direction, or the written value (0/1) changes.
\textbf{(O9)} RowHamemr occurs at both types of cell access transistors, \ie, neighboring and passing gates (\S\ref{subsec:2_2_dram_internal_structure}).
\textbf{(O10)} A victim cell is only susceptible from RowHammer to only one type of the gate at a time, which is reversed when the written value changes.
\textbf{(O11)} Given a particular victim cell, its horizontally adjacent four victim cells' data affect its RowHammer vulnerability, which becomes strongest at a distance of two.
\textbf{(O12)} Similar horizontal influence exists in the aggressor row, which becomes weakest at distance two.
We also find that a newly-discovered adversarial data pattern \textbf{(O13)} decreases the activation count that triggers the first bitflips (\hcnt) in a victim row by up to 81\% and \textbf{(O14)} exacerbates the overall bit error rate (BER) of the victim row by up to 1.69$\times$.
%
%
%
%

\begin{figure}[!tb]
    \center
    \vspace{0.0in}
    \includegraphics[width=0.88\linewidth]{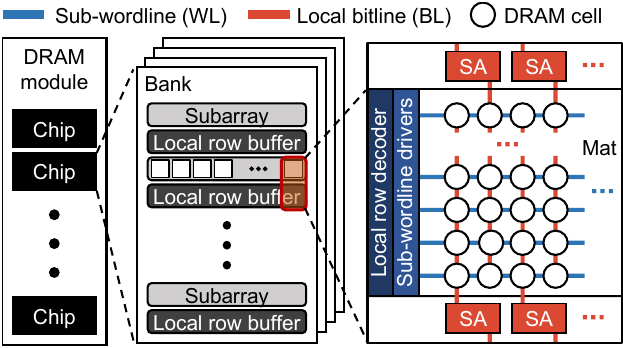}
    \vspace{-0.11in}
    \caption{
        Conventional DRAM organization.
    }
    \vspace{-0.14in}
    \label{fig:DRAM_organization}
\end{figure}

%
%
%
%
%

\noindent \textbf{New vulnerabilities and protection patches (\S\ref{sec:6_vuln_prot}):}
Based on the new observations at both macroscopic and microscopic levels, we identify previously unknown \aib vulnerabilities and propose a simple yet effective protection solution.
First, we identify that edge subarrays and coupled rows can pose a new threat to existing \aib protection solutions, those driven by the memory controller.
Besides, it can facilitate the attacker to succeed in a memory templating/massaging~\cite{sp-2020-rambleed} phase, which is an essential part of a successful attack.
%
%
%
%
%
Second, we demonstrate that, depending on the intention of the \aib attack, we can exploit an adversarial data pattern: (1) to achieve a specific cell's bitflip or (2) to maximize the number of bitflips in the target row by collocating the row and column data/directional dependence. 
%
%
Lastly, countering this, we propose a simple yet effective data masking mechanism that can prevent the exploitation of such vulnerability.

\section{Background}
\label{sec:2_background}

In this section, we present the background on DRAM organization, microarchitectures, and operations. 
We also overview DRAM activate-induced bitflips (\aibs).

\subsection{DRAM Organization}
\label{subsec:2_1_dram_organization}

A DRAM module is hierarchically organized, from top to bottom, chips, banks, subarrays, MATs, and cells (see Figure~\ref{fig:DRAM_organization}).
A DRAM cell consists of a capacitor and an access transistor, indexed by row/column address via the corresponding wordline (WL) and bitline (BL), respectively.
A row decoder enables a specific WL, which turns on the access transistors that connect the cell capacitors to the sense amplifiers (SAs) via BLs.
Each cell stores 1-bit data, and is classified into true-cell (anti-cell) if a charged state represents 1 (0)~\cite{isca-2013-retention,asplos-2019-cta, isca-2024-hifidram}.
An SA senses and amplifies a voltage small difference in a pair of BLs and temporarily stores the value of a cell. 

DRAM cells form a 2D array structure referred to as MAT, and an array of MATs constitute a single subarray.
A single read/write command reads and writes data from one or more MATs.
A subarray can have either an open or folded bitline structure, depending on whether a single SA is connected to both the upper and lower BLs (open) or not (folded)~\cite{2010-DRAM_book}.
In an open BL structure, half BLs of a subarray share SAs with those of the \emph{upper} subarray and the other half BLs of a subarray with those of the \emph{lower} subarray.

\subsection{\sixf Cell Structure}
\label{subsec:2_2_dram_internal_structure}

\begin{figure}[!tb]
  \center
  \vspace{0.0in}
  \includegraphics[width=0.95\linewidth]{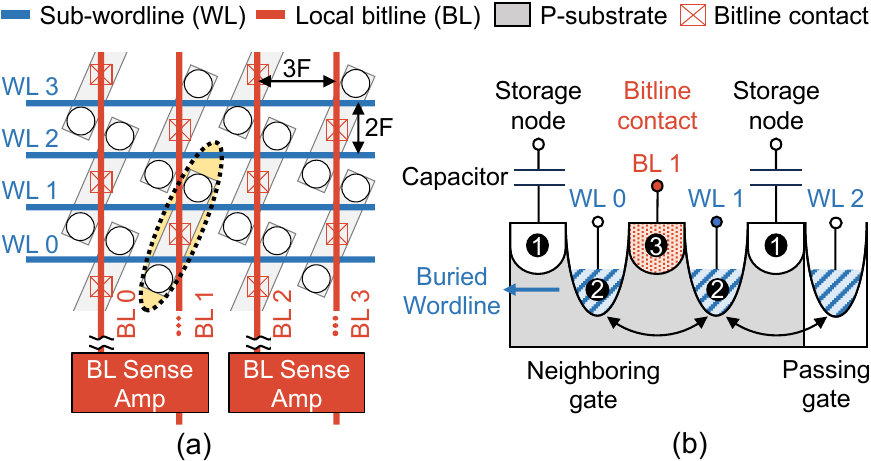}
  \vspace{-0.05in}
  \caption{
    (a) DRAM \sixf cell structure and (b) the cross-section schematic of a saddle-fin transistor.
  }
  \vspace{-0.0in}
  \label{fig:microscopic}
\end{figure}

Modern DRAM chips are primarily designed using a \sixf cell structure~\cite{arxiv-2023-dsac} for higher cell density, where F represents one-half of the minimum pitch (see Figure~\ref{fig:microscopic}(a)).
Figure~\ref{fig:microscopic}(b) shows a cross-section view of the \sixf cell structure's highlighted region in Figure~\ref{fig:microscopic}(a).
The \sixf cell structure adopts a saddle-fin transistor, where a P-type substrate houses {\small\Circled{1}} a pair of DRAM cells, each having a storage node and {\small\Circled{2}} controlled by a buried WLs ($WL0$ or $WL1$). 
{\small\Circled{3}} These two cells share a BL connected through a bitline contact (BC).
When $WL1$ is enabled, we denote the adjacent WL ($WL2$) (which does not share the P-substrate with $WL1$) and the other adjacent WL ($WL0$) as the passing gate and the neighboring gate, respectively~\cite{vlsi-2006-passing_gate, korea-2010-neighboring, arxiv-2023-dsac}.

\subsection{DRAM Operation}
\label{subsec:2_3_dram_operation}

To access data, the memory controller (MC) sends an activate (\texttt{ACT}) command to enable a WL and connect the corresponding row of DRAM cells to BLs.
%
The SAs and BLs are initially precharged to $V_{dd}/2$.
When cells are connected to BLs, charge sharing occurs.
This causes a small deviation in the voltage level of the BL, which is amplified to $V_{dd}$ or 0 by SA.
During the activation of a DRAM row, the global row decoder selects a subarray and the local row decoder selects and drives the corresponding row and WL.
When DRAM receives a read (\texttt{RD}) or write (\texttt{WR}) command, the sensed or to-be-written data pass through the local and global I/O, equipped with temporary buffers on its path (\eg, global dataline SA).

\texttt{tRCD} is the minimum time between an \texttt{ACT} command to the \texttt{RD}/\texttt{WR} command.
The \texttt{RD}/\texttt{WR} command reads/writes data from/to the sensed row in the SAs.
After completing read or write operations, the host sends a precharge (\texttt{PRE}) command to disable the activated row's WL, disconnecting cells from the BLs.
Prior to the precharge, the voltage level of the cell must be restored to $V_{dd}$ or 0.
The required time to issue the \texttt{PRE} command after the \texttt{ACT} command is \texttt{tRAS}. 
After a \texttt{PRE} command is issued, the SAs and BLs require \texttt{tRP} time to restore the BL voltage to $V_{dd}/2$.

\begin{figure}[!tb]
  \center
  \vspace{0.0in}
  \includegraphics[width=0.95\linewidth]{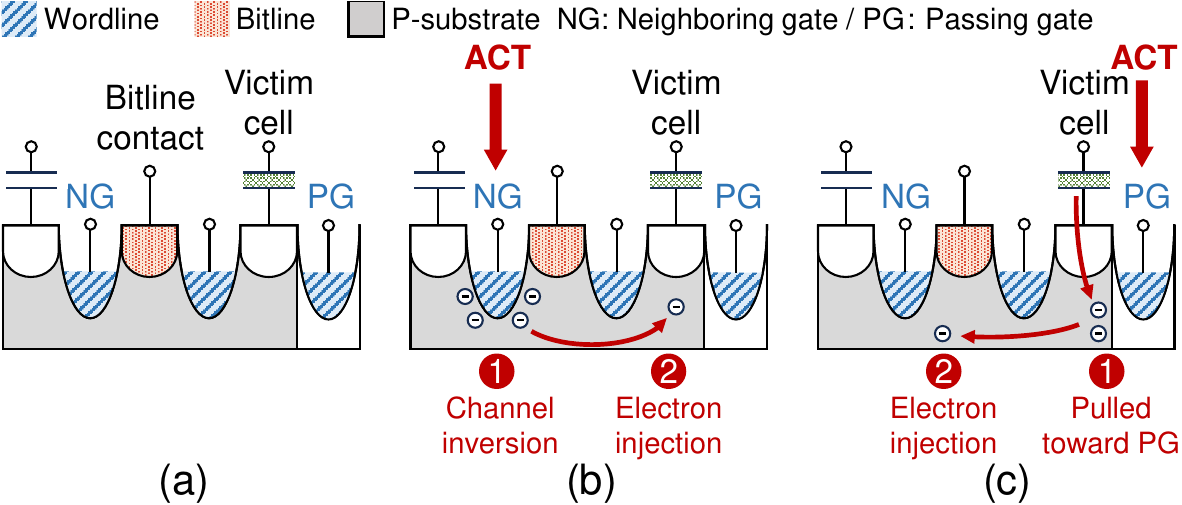}
  \vspace{-0.05in}
  \caption{Mechanisms of activate-induced bitflips. PG and NG denote the passing gate and the neighboring gate, respectively.}
  \vspace{-0.05in}
  \label{fig:bitflips}
\end{figure}

\subsection{DRAM Activate-Induced Bitflips (AIBs)}
\label{subsec:2_4_activate-induced_bitflips}

\aibs are the representative DRAM read disturbance errors wherein activation disturbs cells in adjacent rows and flips the states of the cells (see Figure~\ref{fig:bitflips}).
%
There are two mechnisms occuring AIBs: (1) electron migration (injection and capture) or (2) capacitive crosstalk~\cite{ted-2020-passing-gate,ted-2021-physics,iedm-2017-reliability,irps-2023-double-sided}.
Based on bitflip mechanisms, the pair of an aggressor's WL and a victim cell is decided.
The value of a victim cell can be flipped by electron migration from hammering the neighboring gate or by capacitive crosstalk from hammering the passing gate.
%
Also, depending on the access pattern, \aibs can be caused by two attack patterns: (1) repetitive activation of the specific/aggressor row (RowHammer) or (2) activating the specific/aggressor row over an extended period (RowPress)~\cite{isca-2014-flipping, isca-2023-rowpress, arxiv-2023-dsac}.
%

%
The mechanism of \aibs depends on what type of gate the aggressor's WL is with the victim cell.
Figure~\ref{fig:bitflips}(b) demonstrates \aib by electon injection when the neighboring gate is turned on.
When the aggressor's WL (the neighboring gate) is activated, {\small\Circled{1}} electrons accumulate around the buried WL due to channel inversion.
Upon deactivation, {\small\Circled{2}} the accumulated electrons are spread out, while some are injected into the victim cell sharing the P-substrate. 
Figure~\ref{fig:bitflips}(c) illustrates \aib by electron spreading originating from the passing gate.
When the aggressor’s WL (the passing gate) is activated, {\small\Circled{1}} electrons are continuously attracted from the victim cell toward the passing gate. 
After the row is precharged, {\small\Circled{2}} the electrons are spread out and some are injected into the active region, instead of returning to the victim cell. 
As both are the processes of victim cells acquiring or losing electrons, their likelihoods are affected by the data written to the victim cells~\cite{iedm-2017-reliability,ted-2020-passing-gate}.

RowHammer and RowPress have different access patterns and bitflip characteristics~\cite{arxiv-2023-dsac,isca-2023-rowpress, isca-2014-flipping, isca-2020-revisit}.
RowHammer, which has an attack pattern of repeatedly activating and precharging a single row, can unintentionally affect cell values in its physically-adjacent rows.
RowPress, which keeps a single row activated for a long time, can cause errors in its physically nearby rows with a much lower activation count (the number of ACT-PRE command pairs applied to the row in the intervals of its adjacent rows being refreshed).
Unlike RowHammer, which causes bitflips regardless of the cell's charge, RowPress specifically induces bitflips only in the charged state~\cite{isca-2023-rowpress}.
Accordingly, the factors affecting the \aib phenomenon can be categorized into four specific types: 1) the attack patterns (RowHammer vs. RowPress), and 2) the type of gates (the neighboring gate vs. the passing gate).

\section{Experimental Methodology}
\label{sec:3_experiments}

We first introduce the experimental setup and three utilized reverse-engineering techniques: \aib, RowCopy, and retention-time test.
We point out common pitfalls in reverse-engineering procedures that often stem from complexities in physical to DRAM address mapping~\cite{isca-2010-rethinking-dram} and twisted data pin connection to each chip.
Such pitfalls can be avoided based on publicly available information, which is yet scattered across various
documents.

\subsection{FPGA-based Testing Infrastructure}
\label{sec:3-1_fpga_environment}

\begin{figure}[!tb]
  \center
  \vspace{0.0in}
  \includegraphics[width=0.9\linewidth]{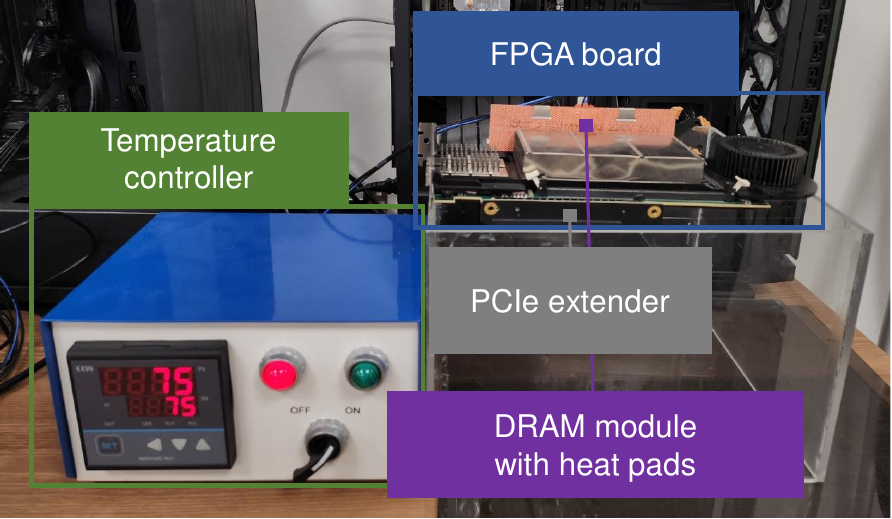}
  \vspace{-0.05in}
  \caption{
    Test infrastructure for DDR4 DIMM and HBM2.
  }
  \vspace{-0.15in}
  \label{fig:fpga}
\end{figure}

We modified SoftMC~\cite{hpca-2017-softmc} and DRAM Bender~\cite{tcad-2023-dram-bender} to execute the three DRAM reverse-engineering techniques and setup an FPGA-based DRAM testing platform (Figure~\ref{fig:fpga}).
We constructed our FPGA-based DRAM testing platform using Xilinx Alveo U200~\cite{u200} and U280~\cite{u280} for testing DDR4 and HBM2, respectively.
%
We tested $376$ DDR4 chips from three major DRAM manufacturers (160 chips from \samsung, 128 chips from \hynix, and 88 chips from \micron), and $4$ HBM2 stacks from \samsung~\cite{xilinx-hbm2} for our experiments (\cf Table~\ref{tbl:3_tested_chips} for more details).
Additionally, we presume no aging effect in DRAM devices' fault rate according to the previous works~\cite{dft-2017-lifetime,dsn-2015-aging_meta, sc-2013-feng_shui}.
Therefore, we conducted our experiments on various DRAM chips from 2016 to 2021 using the same experimental methodology.
We controlled the DRAM testing platform to issue consecutive DRAM commands to DDR4 and HBM2 with a minimum interval of 1.25ns and 1.67ns (equal to tCK), respectively.
%
We also employed a temperature controller and silicon rubber heaters to control the temperature of DRAM chips.
We performed our experiments with DDR4 DIMMs at 75$^\circ$C,\footnote{Although RowHammer~\cite{isca-2014-flipping,isca-2020-revisit,micro-2021-deeper,arxiv-2023-spyhammer} and RowPress~\cite{isca-2023-rowpress} are both known to exhibit temperature-dependent error behaviors, 
we did not observe significant 
differences in trends at other temperatures, which did not change our key observations and conclusions.} whereas we tested HBM2 at a constant room temperature as we could not regulate the temperature of HBM2.

\begin{table}[]
    \centering
    \caption {The tested DDR4 and HBM2 chips.}
    \begin{tabular}{cccccc}
        \Xhline{2\arrayrulewidth}
        \textbf{DRAM type} & \textbf{Vendor} & \textbf{Chip type} & \textbf{Density} & \textbf{Year} & \textbf{\# chips} \\ \Xhline{2\arrayrulewidth}
        \multicolumn{1}{c}{\multirow{6}{*}{DDR4}} 
        
                                    & \multirow{6}{*}{\samsung}      
                                               & $\times$4          & 8Gb       & 2016   & 80\\
        \multicolumn{1}{c}{}        & {}       & $\times$4          & 8Gb       & 2017   & 16\\
        \multicolumn{1}{c}{}        & {}       & $\times$4          & 8Gb       & 2018   & 32\\
        \multicolumn{1}{c}{}        & {}       & $\times$4          & 8Gb       & 2021   & 32\\ 
        \multicolumn{1}{c}{}        & {}       & $\times$8          & 8Gb       & 2017   & 16\\ 
        \multicolumn{1}{c}{}        & {}       & $\times$8          & 8Gb       & 2018   & 32\\ 
        \multicolumn{1}{c}{}        & {}       & $\times$8          & 8Gb       & 2019   & 16\\ 
        \midrule 
        
        \multicolumn{1}{c}{\multirow{4}{*}{DDR4}}   & \multirow{4}{*}{\hynix}        
                                               & $\times$4          & 8Gb       & 2019   & 64\\ 
        \multicolumn{1}{c}{}        & {}       & $\times$8          & 8Gb       & 2017   & 32\\
        \multicolumn{1}{c}{}        & {}       & $\times$8          & 8Gb       & 2018   & 24\\
        \multicolumn{1}{c}{}        & {}       & $\times$8          & 8Gb       & 2019   & 8\\
        \midrule 
        
        \multicolumn{1}{c}{\multirow{4}{*}{DDR4}}   & \multirow{4}{*}{\micron}        
                                               & $\times$4          & 8Gb       & 2018   & 32\\
        \multicolumn{1}{c}{}        & {}       & $\times$4          & 8Gb       & 2021   & 32\\
        \multicolumn{1}{c}{}        & {}       & $\times$8          & 8Gb       & 2016   & 8\\
        \multicolumn{1}{c}{}        & {}       & $\times$8          & 8Gb       & 2019   & 16\\
        \midrule 
        
        \multicolumn{1}{c}{HBM2}    & \samsung      & 4-Hi stack & 4GB/stack & N/A     & 4\\ 
        \bottomrule
        \label{tbl:3_tested_chips}
    \end{tabular}
    \vspace{-0.3in}
\end{table}

\subsection{DRAM Reverse-engineering Techniques}
\label{sec:3-2_reverse_techniques}
We utilize the three DRAM reverse-engineering techniques to uncover the DRAM microarchitecture and operations, and analyze the DRAM \aib characteristics.

\noindent
\textbf{Activate-induced bitflips (\aibs)} can indicate which row is adjacent to the activated aggressor row based on the fact that the physically most adjacent row is affected most~\cite{isca-2020-revisit,ted-2021-physics}.
Most rows have two physically adjacent rows (above and below).
However, the row at the edge of a subarray boundary has only one physically adjacent row.
%
The causes of \aibs are the injection/removal of electrons into/from the victim cell and capacitive crosstalk, depending on whether the WL of the aggressor row is a neighboring gate or a passing gate (\S\ref{subsec:2_4_activate-induced_bitflips}).
%

\noindent
\textbf{RowCopy}~\cite{micro-2013-rowclone} is an out-of-specification in-memory operation that copies the value of one row to another row within the same subarray using charge-sharing between a BL and a cell. 
First, a source row is activated. After tRAS, the row is precharged. 
However, if the destination row is activated soon enough, the BL will not be fully precharged to $V_{dd}/2$ yet. 
Because the capacitance of a BL is much larger than a cell, the source row values can be effectively copied to the destination row through a charge transfer from the BL to the cell. 
We identify the height of the subarrays because RowCopy is not possible between other subarrays.
Also, exploiting the fact that adjacent subarrays share half of the SAs in an open bitline structure, we identify the type of subarray structure (open or folded bitline) for each tested DRAM.
%

\noindent
\textbf{Retention time test}~\cite{isca-2013-retention} allows us to correctly distinguish between true-cells and anti-cells.
Value 1 is represented by charged and discharged states in true-cells and anti-cells, respectively.
This is a design choice to reduce the noise or optimizing the data path from the SAs to the I/O~\cite{isca-2013-retention}.
The DRAM cells naturally leak charge over time, which leads to retention failure unless periodically refreshed.
The retention time of a cell is the length of time before it loses its data.
Exploiting the fact that leakage occurs from a charged state to a discharged state, we perform a retention time test to distinguish between true-cells and anti-cells.
We discovered that only true-cells are used in \samsung and \hynix's DRAM chips, whereas both the true-cells and anti-cells are interleaved at a subarray granularity in \micron's DRAM chips.

\begin{figure}[!tb]
  \center
  \vspace{0.0in}
  \includegraphics[width=0.92\linewidth]{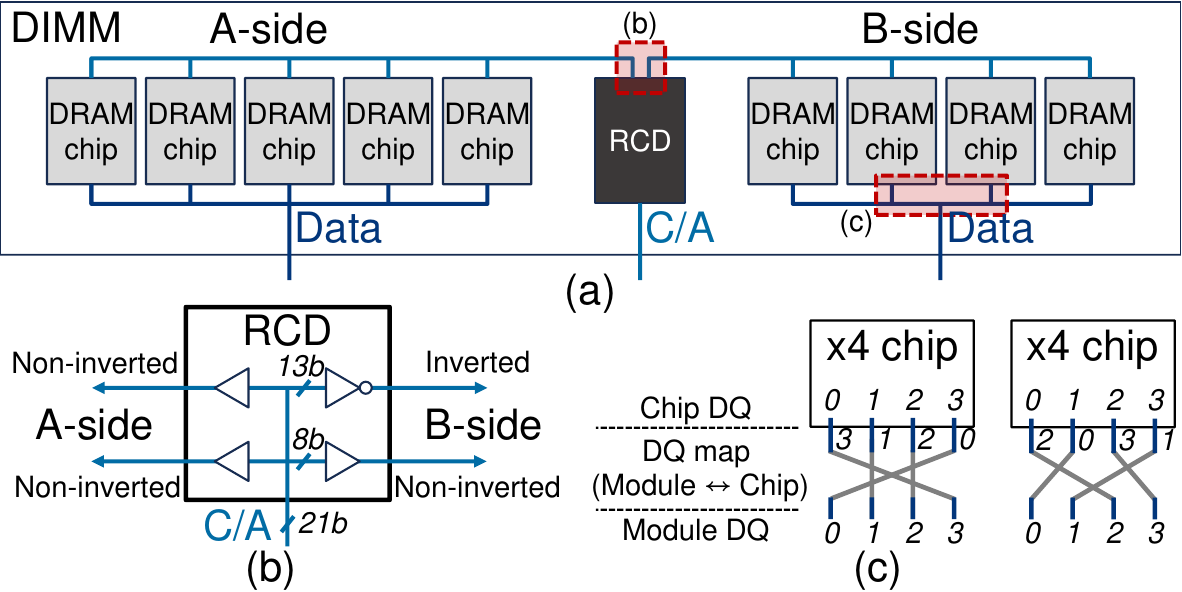}
  \vspace{-0.11in}
  \caption{Identified DRAM mapping for reverse-engineering.}
  \label{fig:dram_mapping}
  \vspace{-0.12in}
\end{figure}

\subsection{Common Pitfalls from Address and Data Mapping}
\label{sec:3-3_common_pitfalls}

While complexities in physical-to-DRAM address mapping and data pin (DQ) twisting (or remapping) 
are disclosed in publicly available documents, they are regularly overlooked, leading to the following common pitfalls; (1) row address remapping at a registered clock driver (RCD) chip, (2) row address remapping in a DRAM chip, and (3) DQ remapping at each chip.

\noindent
\textbf{Common pitfall-(1):} 
The row address can be remapped at the RCD chip~\cite{jedec-ddr4-rcd}.
The RCD chip of the registered DIMM (RDIMM) or load-reduced DIMM (LRDIMM) reduces the MC's driving load to broadcast command/address (C/A) signals to multiple DRAM chips by decoupling the the C/A signals driven by the MC from all the DRAM chips (Figure~\ref{fig:dram_mapping}(a)).
As illustrated in Figure~\ref{fig:dram_mapping}(b), \emph{address inversion} is enabled by default for the RCD chip to conserve power and reduce simultaneous output switching current~\cite{jedec-ddr4-rcd}.
When the address inversion is turned on, some of the row and bank addresses to the DRAM chips on B-side (right in Figure~\ref{fig:dram_mapping}(b)) are inverted, while A-side (left in Figure~\ref{fig:dram_mapping}(b)) receives the non-inverted address.
Such inversion can be easily neglected, which can lead to misinterpreted observations, such as direct non-adjacent RowHammer effect\footnote{A phenomenon where frequently activating $N^{th}$ row can \emph{directly} affect not only distance 1 (\ie, $N\!\pm\!1^{th}$ rows) but also distance 3, 5, and further away rows.}~\cite{isca-2020-revisit}, half-row~\cite{isca-2014-half-dram}, and incorrectly interpreted spare rows~\cite{sp-2020-suscep, hpca-2015-cidra}.
When the inversion was considered, we were not able to observe the three phenomenons (concurring with prior study~\cite{msr-2021-mfit}), whereas we could reproduce them when we disregarded such inversion.
We took the inversion fully into account in our analysis, similar to prior studies~\cite{msr-2021-mfit}.

\noindent
\textbf{Common pitfall-(2):}
The row address can also be remapped by the internal remapping scheme of each DRAM chip.
For example, while DRAM decoders may preserve the sequential row order when mapping from physical addresses to DRAM row addresses, they can also scramble the row order.
%
%
In a similar way to prior studies~\cite{isca-2020-revisit, micro-2021-deeper, isca-2014-flipping}, we reconstructed internal row mapping by executing single-sided RowHammer attacks.
The two rows with the most errors are the physically most adjacent rows.
%
We found that only DDR4 and HBM2 of \samsung remapped rows internally, while DDR4 of \hynix and \micron did not.
%
%
From now on, we base our analysis on the remapped row addresses.

\noindent
\textbf{Common pitfall-(3):}
Most prior studies~\cite{isca-2014-flipping,sp-2020-suscep,isca-2020-revisit,micro-2021-uncovering} commonly use data patterns, such as $\mathtt{0x55}$ or $\mathtt{0xAA}$ assuming a straightforward connection.
However, we note that DQ pins from a DIMM to each DRAM chip are also remapped (Figure~\ref{fig:dram_mapping}(c))~\cite{jedec-ddr4-rdimm, micron-dimm16GB, micron-dimm32GB}.
%
%
Therefore, even though the user writes $\mathtt{0x55}$, each DRAM chip can receive different data (\eg, $\mathtt{0x33}$, $\mathtt{0xCC}$, or $\mathtt{0x99}$).
To write the same data into all DRAM chips, we thoroughly took DQ twisting into consideration.

\begin{table}
    \centering
    \small
    \vspace{-0.05in}
    \caption{Terminologies used throughout \S\ref{sec:4_contribution1}}
    \vspace{-0.1in}
    \label{tbl:4_contribution1}
    \resizebox{1\columnwidth}{!}{
    \begin{tabular}{p{0.25\columnwidth} p{0.67\columnwidth}} 
        \Xhline{3\arrayrulewidth}
        \textbf{Symbol} & \textbf{Description} \\ 
        \Xhline{1.5\arrayrulewidth}
        Data swizzling  & The data reordering that occurs when transferred from MC to DRAM, and vice-versa.  \\
        \rdgranul & The amount of data that is read from a single chip for a single RD command (\eg, 32-bit). \ie, cache-line width divided by the number of chips. \\
        Edge subarray   & The subarray that is at the physical edge, with only one neighboring subarray. \\
        \emph{Even/odd BL}     & BL that is indexed by an even/odd number. \\
        \Xhline{3\arrayrulewidth}
    \end{tabular}
    }
\end{table}

\section{Macroscopic Analysis of DRAM Microarchitecture}
\label{sec:4_contribution1}

We conduct a \emph{macroscopic} analysis on DRAM microarchitecture. 
First, we reverse-engineer the \emph{data swizzling} that occurs between MC to DRAM, which is utilized across the rest of the analysis.
Then, multiple observations are presented in the order of MAT, row, and subarray.
Table~\ref{tbl:4_contribution1} summarizes the terminologies that are frequently used in this section.
%
%
%

\begin{figure}[!tb]
  \center
  \vspace{0in}
  \includegraphics[width=0.88\columnwidth]{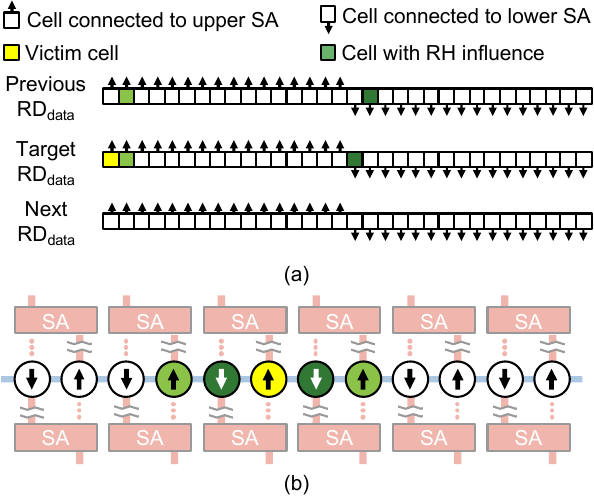}
  \vspace{-0.12in}
  \caption{
    (a) An example of reverse-engineering the data swizzling and (b) its cell layout view with open BL structure.
    We identify the most influenced cells to each cell (victim cell) for RowHammer and distinguish whether they are connected to an upper SA or a lower SA by RowCopy.
  }
  \vspace{-0.12in}
  \label{fig:column_address_reverse_engineering}
\end{figure}

\subsection{Data Swizzling and MAT Structure}
\label{sec:4-1_macro_mat}

We reverse-engineered the data swizzling based on \aib and identified the MAT width.
%
We faithfully considered DQ twisting, the horizontal \aib influence \textbf{(O11)}, and the \emph{even/odd BL} distinguished by RowCopy.
We define the \emph{even/odd BL} as the BL indexed by an even/odd number when we set an ordered number starting from the physically leftmost BL.
The horizontal influence, or the fact that physically adjacent \emph{victim} cells affect \aib, is elaborated later (\S\ref{sec:5_contribution2}).
%

First, based on the observed fact that horizontally adjacent cells impact \aib, we find the set of cells that are adjacent to a certain victim cell, in a brute-force way.
%
Figure~\ref{fig:column_address_reverse_engineering}(a) illustrates our testing methodology.
For each specific victim cell, we identified four different victim cells with the largest influence.
Some were in the same \rdgranul while others were in the adjacent \rdgranul.
We define the \rdgranul as the data being read from a single \emph{chip} for a single RD command (\eg, 32-bit for a $\times$4 chip).
For example, from \samsung DDR4 $\times$4 chips, we observed that bit 0 of an \rdgranul is influenced by bit 1 and 16 of the same \rdgranul, and bit 1 and 17 of the previous \rdgranul.
Repeated experiments granted us the set of horizontally adjacent cells.
However, because the distance $\pm1$ and $\pm2$ cells have indistinguishable differences in influence (\textbf{O11}), full mapping could not be acquired.

\begin{figure}[!tb]
  \center
  \vspace{0.0in}
  \includegraphics[width=0.9\columnwidth]{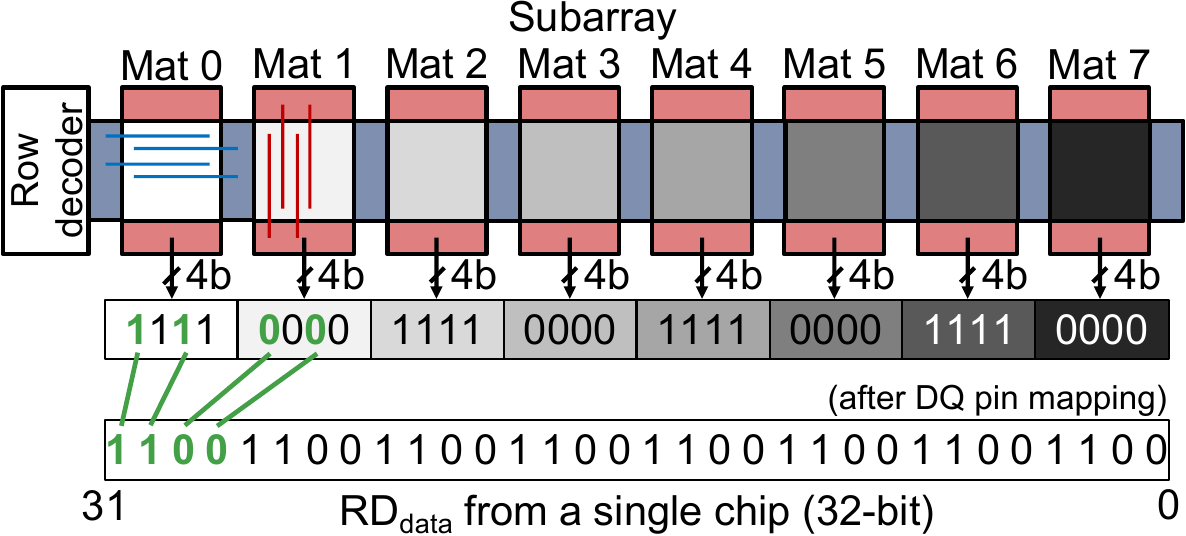}
  \vspace{-0.1in}
  \caption{
    The specific manner in which 32-bit \rdgranul is physically stored in a \samsung's DDR4 $\times$4 chip.
  }
  \vspace{-0.15in}
  \label{fig:mat}
\end{figure}

Second, we utilized RowCopy to distinguish the \emph{odd} and \emph{even BL}, allowing us to distinguish distance $\pm$1 and $\pm$2 cells.
As explained before, in the open bitline structure, half of the cells (\eg, odd) are connected to the upper SAs, while the other half (\eg, even) are linked to the lower SA.
Exploiting this, we discover if each of the collected adjacent cells have \emph{even} or \emph{odd BL} (Figure~\ref{fig:column_address_reverse_engineering}(a)).
When we sequentially examine each cell, checking its four adjacent cells and which has odd/even BL, we can gain the full data swizzling as shown in Figure~\ref{fig:column_address_reverse_engineering}(b).
%

Lastly, while we identified cells in a row that influence each other, we also recognized a set of cells that are isolated from each other. 
%
%
%
We speculate such isolation is due to peripheral circuits between the MATs, such as the local row decoder and sub-WL drivers (see Figure~\ref{fig:mat}).
This makes it difficult for a cell in one MAT to affect a cell in another MAT.
Thus, we group such isolated cells while sweeping every \rdgranul, which indicates the width of a MAT. 
%
%
%
In the tested $\times$4 DDR4 chips, the measured MAT widths are 512-bit, 1024-bit, and 512-bit for \samsung, \hynix, and \micron, respectively.

Based on these processes, we could reconstruct the final data swizzling as shown in Figure~\ref{fig:mat}.
%
The 32-bit \rdgranul of \samsung's $\times$4 DRAM chip is collected from 8 different MATs, each provisioning 4-bit.
Each 4-bit from a MAT is again reorganized, as shown in the green line of the figure.
However, while we numbered each MAT from 0 to 7 for convenience, we could not figure out the physical ordering of each MAT.
%
%
%
%

\vspace{-0.04in}
\begin{tcolorbox}[boxsep=0pt,left=3pt,right=3pt,arc=0pt]
\emph{\textbf{Observation-1:}}
The data of a single RD command is collected from multiple MATs and reorganized due to data swizzling.
\end{tcolorbox}
\vspace{-0.05in}

\vspace{-0.04in}
\begin{tcolorbox}[boxsep=0pt,left=3pt,right=3pt,arc=0pt]
\emph{\textbf{Observation-2:}}
The MAT width, or the number of cells in a row within a single MAT, is measured to be 512- or 1024-bit for tested $\times$4 DDR4 chips.
\end{tcolorbox}
\vspace{-0.05in}

\begin{figure}[!tb]
  \center
  \vspace{0in}
  \includegraphics[width=1\linewidth]{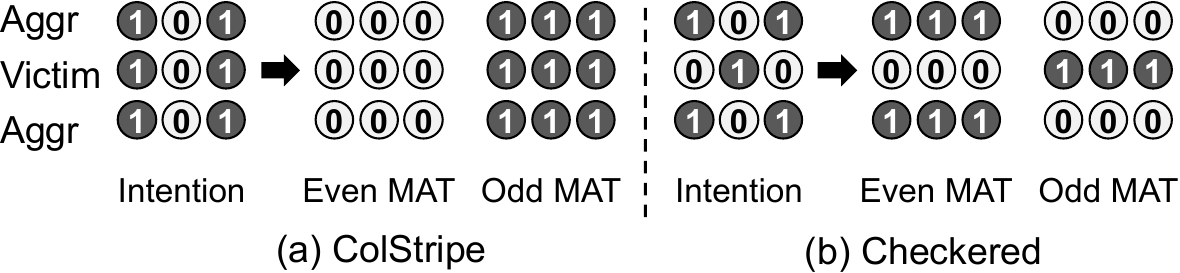}
  \vspace{-0.2in}
  \caption{
  Unintended arrangement of data in even and odd MATs for commonly used data patterns (ColStripe and Checkered) without considering DRAM internal column address mapping.
  }
  \vspace{-0.1in}
  \label{fig:prior_data_pattern}
\end{figure}

Our newly found data swizzling also suggests that the impact of previously understood data patterns on \aib is imprecise.
For example, a `ColStripe' pattern~\cite{delta-2002-datapattern,isca-2020-revisit,sp-2020-suscep,micro-2021-deeper}, which alternates the data for every BL, actually acts as a `Solid' pattern without proper mapping (Figure~\ref{fig:prior_data_pattern}(a)).
Similarly, a `Checkered' pattern acts as a `RowStripe' pattern (Figure~\ref{fig:prior_data_pattern}(b)).
Our detailed analysis on the data patterns in association with the newly discovered mapping and \sixf is provided in \S\ref{sec:5_contribution2}.

\subsection{Coupled-row Activation}
\label{sec:4-2_macro_row}

A coupled-row activation is identified for certain $\times$4 DRAM chips. 
Both RowCopy and \aib indicate that when a row is activated (\eg, $i$th row), its coupled row (\eg, ($i+N_{row}/2$)th row) is activated as well.
We dub such two rows indexed by two different rows specified by physical address but mapped to the same single DRAM row as a coupled-row pair.
Such a behavior was exhibited by \samsung and \hynix's $\times$4 DDR4 chips and \samsung's HBM2.
We speculate that this is a result of an optimization that ensures multiple DRAM types (\eg, $\times$4 and $\times$8) to maintain the same density of cells per WL (\eg, 4096-bit or 8192-bit), regardless of the DRAM I/O width.
%
%
%
This can serve as another \aib vulnerability, unless the host is aware of this coupling and applies proper mitigation to both the victim row and its coupled row.

\vspace{-0.04in}
\begin{tcolorbox}[boxsep=0pt,left=3pt,right=3pt,arc=0pt]
\emph{\textbf{Observation-3:}}
For some DRAM chips, activating a row can result in the unintended activation of the coupled row.
\end{tcolorbox}
\vspace{-0.05in}

\subsection{Subarray Structure}
\label{sec:4-3_macro_subarray}

The subarray height of a DRAM chip can be accurately verified using the RowCopy-based reverse-engineering technique, and cross-checked with \aibs.
%
%
Half of the cells share SAs with the upper or lower subarrays in an open bitline structure (\S\ref{subsec:2_1_dram_organization}).
%
%
Thus, when we look for row address boundaries where RowCopy starts to work only for half of the cells, we can identify (1) the subarray boundary and (2) which two subarrays are adjacent.
Through experiments, we have discovered that all tested chips have an open bitline structure.\footnote{Depending on the manufacturer, generation, and type, the exact bit/column location where only half of the row succeeds in RowCopy differs. This could be due to variations in the column decoder or connection between local and global I/O.}
Also, while \samsung and \hynix copied the data in an inverted form due to the SA structure, \micron copied the data as is because \micron has true-/anti-cell interleaved at the subarray granularity (\S\ref{sec:3-2_reverse_techniques}).

\aib can also be utilized to identify the subarray heights~\cite{hotos-2021-stop,msr-2021-mfit,sosp-2023-siloz}.
%
SAs that are more than 100 times larger than a DRAM cell~\cite{rambus-dram,isca-2024-hifidram} separate two different subarrays, preventing \aib from occurring between two rows that are physically separated by SAs.
Consequently, only the aggressor row belonging to the same subarray and adjacent to the victim row can cause bitflips.
Prior studies also exploited RowHammer to find subarray heights~\cite{msr-2021-mfit, arxiv-2023-hbm2_rowhammer} or used the DRAM command sequence (\texttt{ACT-PRE-ACT}) to derive the adjacency between two subarrays~\cite{micro-2022-hira}.
However, because RowCopy can provide information on both aspects and is more efficient in time than the \aib method, we mainly used RowCopy and relied on \aib only for validation.

\begin{table}[]
    \centering
    \caption{The structure of subarrays and rows we identified.}
    \label{tbl:4_subarray_size}
    \resizebox{1\columnwidth}{!}{
        \begin{tabular}{lllllll}
        \toprule
        \textbf{\begin{tabular}[c]{@{}c@{}}DRAM\\ type\end{tabular}}
        & \textbf{Vendor} & \textbf{\begin{tabular}[c]{@{}c@{}}Chip\\ type\end{tabular}}
        & \textbf{Year} & \textbf{\begin{tabular}[c]{@{}c@{}}Subarray\\ composition\end{tabular}}  
        & \textbf{\begin{tabular}[c]{@{}c@{}}Edge \\subarray\\interval\end{tabular}} 
        & \textbf{\begin{tabular}[c]{@{}c@{}}Coupled- \\row\\ distance\end{tabular}} \\
        \toprule
        \multirow{26}{*}{DDR4} 
            & \multirow{10}{*}{\samsung} 
                & \multirow{4}{*}{$\times$4} 
                    & \begin{tabular}[c]{@{}c@{}} 2016\\2017\end{tabular}
                        & \begin{tabular}[c]{@{}c@{}} 11 $\times$ 640-row\\ 2 $\times$ 576-row\\(per 8192-row)\end{tabular}
                            & \begin{tabular}[c]{@{}c@{}} per\\16K rows\end{tabular}   
                                & 64K rows 
                                    \\ \cline{4-7}
            &   &   & \begin{tabular}[c]{@{}c@{}} 2018\\2021\end{tabular}
                        & \begin{tabular}[c]{@{}c@{}} 4 $\times$ 832-row\\ 1 $\times$ 768-row\\(per 4096-row)\end{tabular}
                            & \begin{tabular}[c]{@{}c@{}} per\\32K rows\end{tabular}  
                                & N/A  
                                    \\ \cline{3-7} 
            &   &   & \begin{tabular}[c]{@{}c@{}} 2017\\2019\end{tabular}
                        & \begin{tabular}[c]{@{}c@{}} 11 $\times$ 640-row\\ 2 $\times$ 576-row\\(per 8192-row)\end{tabular}
                            & \begin{tabular}[c]{@{}c@{}} per\\16K rows\end{tabular}  
                                & N/A  
                                    \\ \cline{4-7}  
                &   & \multirow{-4}{*}{$\times$8} 
                    & \begin{tabular}[c]{@{}c@{}} 2018\end{tabular}
                        & \begin{tabular}[c]{@{}c@{}} 4 $\times$ 832-row\\ 1 $\times$ 768-row\\(per 4096-row)\end{tabular}
                            & \begin{tabular}[c]{@{}c@{}} per\\32K rows\end{tabular}    
                                & N/A 
                                    \\ \cline{2-7}
            & \multirow{4}{*}{\hynix} 
                & $\times$4   
                    & 2019 
                        & \begin{tabular}[c]{@{}c@{}} 4 $\times$ 832-row\\ 1 $\times$ 768-row\\(per 4096-row)\end{tabular} 
                            & \begin{tabular}[c]{@{}c@{}} per\\32K rows\end{tabular}    
                                & 64K rows  
                                    \\ \cline{3-7} 
            &   & $\times$8       
                    & \begin{tabular}[c]{@{}c@{}} 2017\\2018\\2019\end{tabular}
                        & \begin{tabular}[c]{@{}c@{}} 4 $\times$ 832-row\\ 1 $\times$ 768-row\\(per 4096-row)\end{tabular} 
                            & \begin{tabular}[c]{@{}c@{}} per\\32K rows\end{tabular}    
                                & N/A    
                                    \\ \cline{2-7} 
            & \multirow{7}{*}{\micron} 
                & $\times$4   
                    & \begin{tabular}[c]{@{}c@{}} 2018\\2021\end{tabular}
                        & \begin{tabular}[c]{@{}c@{}}2 $\times$ 688-row\\ 1 $\times$ 672-row\\ (per 2048-row)\end{tabular} 
                            & \begin{tabular}[c]{@{}c@{}} per\\32K rows\end{tabular}    
                                & N/A 
                                    \\ \cline{3-7} 
            &   & \multirow{4}{*}{$\times$8} 
                    & 2016
                        & \begin{tabular}[c]{@{}c@{}}1 $\times$ 688-row\\ 2 $\times$ 680-row\\ (per 2048-row)\end{tabular} 
                            & \begin{tabular}[c]{@{}c@{}} per\\4K rows\end{tabular}    
                                & N/A
                                    \\ \cline{4-7}
            &   &   & 2019
                        & \begin{tabular}[c]{@{}c@{}}2 $\times$ 688-row\\ 1 $\times$ 672-row\\ (per 2048-row)\end{tabular} 
                            & \begin{tabular}[c]{@{}c@{}} per\\32K rows\end{tabular}    
                                & N/A \\     
            \midrule
        HBM2& \samsung                  
                & 4-Hi    
                    & N/A    
                        & \begin{tabular}[c]{@{}c@{}} 4 $\times$ 832-row\\ 1 $\times$ 768-row\\(per 4096-row)\end{tabular}    
                            & \begin{tabular}[c]{@{}c@{}} per\\8K rows\end{tabular}     
                                & 8K rows \\    
        \bottomrule
        \end{tabular}
    }
\end{table}

Utilizing the aforementioned methodology, we discovered that the subarray height is not a power of 2 and also can vary within a single chip, in contrast to the common conception.
%
\samsung's DDR4 $\times$4 chips (2016 and 2017) and $\times$8 chips (2017 and 2019) have a repeated pattern of 11 subarrays with 640 rows and two subarrays with 576 rows (a total of 8192 rows).
%
The other DDR4 chips from \samsung have a pattern of four subarrays with 832 rows and one subarray with 768 rows (a total of 4096 rows) is repeated.
We have also identified that the \hynix's DDR4 chips and \samsung's HBM2 chips have identical subarray structures with the up-to-date (till 2021) \samsung's DDR4 DRAM.
By contrast, \micron's DDR4 chips made in 2016 and 2018-2021 have a pattern of one subarrays with 688 rows and two subarray with 680 rows (a total of 2048 rows) and 688 rows and one subarray with 672 rows (a total of 2048 rows), respectively.
%
We presume that the varying height of the subarray is a compromise between deteriorating timing parameters and higher cell density and fewer SAs when the cell per BL (subarray height) increases.
This concurs with the trend that the subarray height has been increasing with the DRAM technology scaling.
Table~\ref{tbl:4_subarray_size} summarizes the discovered subarray compositions.

\vspace{-0.04in}
\begin{tcolorbox}[boxsep=0pt,left=3pt,right=3pt,arc=0pt]
\emph{\textbf{Observation-4:}}
The subarray heights are not power of 2, and different across different generations and within a chip. 
\end{tcolorbox}
\vspace{-0.05in}

\begin{figure}[!tb]
  \center
  \vspace{0.0in}
  \includegraphics[width=1\linewidth]{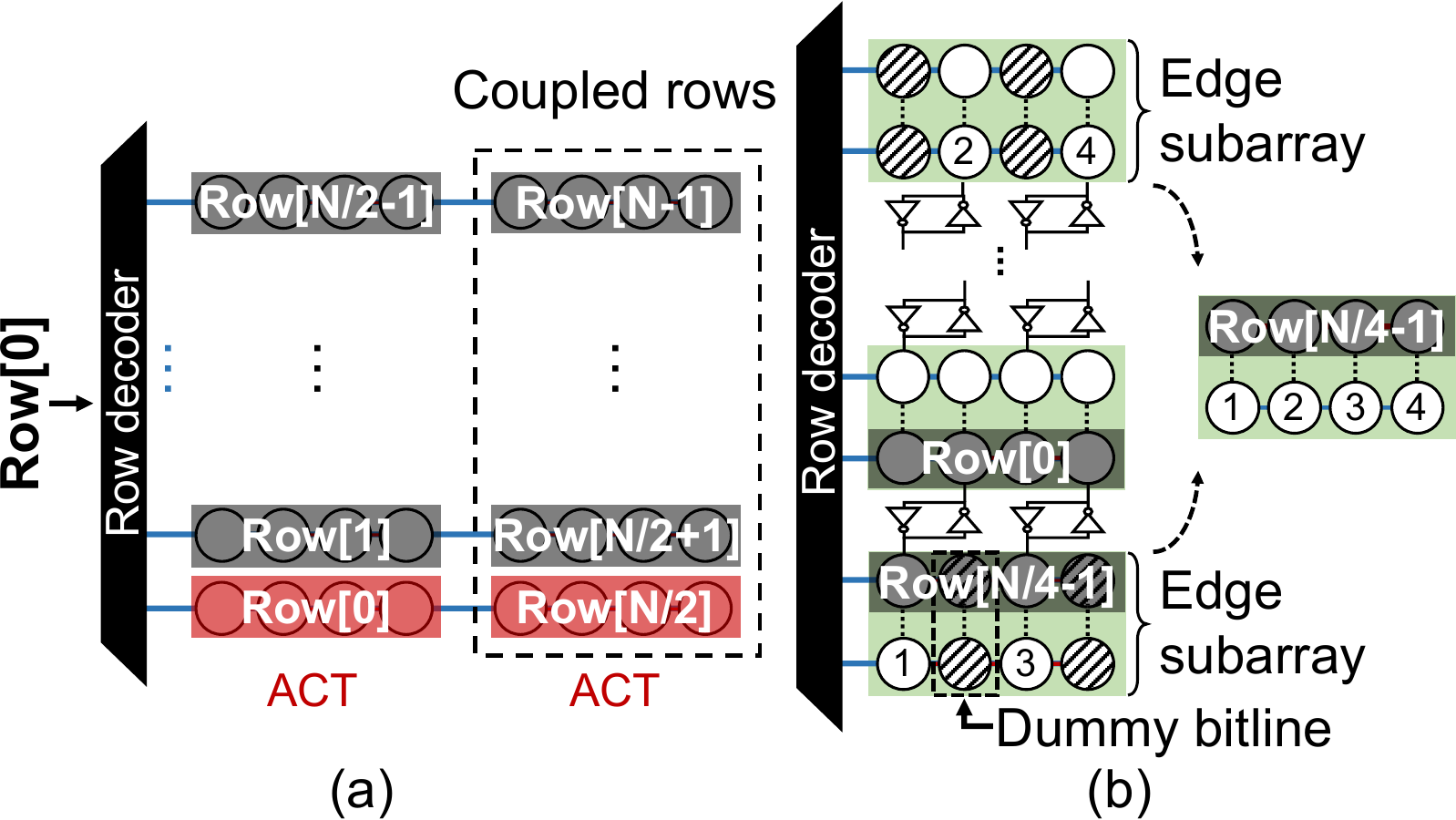}
  \vspace{-0.2in}
  \caption{
    Characteristics of subarray structures: (a) Activating a row could incur activating its coupled row. (b) An edge subarray physically consists of a pair of subarrays, each having dummy bitlines.
  }
  \vspace{-0.1in}
  \label{fig:subarray_structure}
\end{figure}

We also observed that two different edge subarrays, or subarrays that are at the physical edge with only one neighboring subarray, work in tandem to create a single full subarray.
%
%
While we found out that most subarrays are sequentially adjacent, following the row address order, we identified corner cases.
For example, for some tested DRAM chips, when RowCopy was executed for the $0$\textsuperscript{th} row as a source and the ($N_{row}/4 - 1$)\textsuperscript{th} row as a destination, half of the bits were copied despite the large difference in the row address.
$N_{row}$ denotes the total number of rows in a bank (\eg, $2^{17}$).
Because the $0$\textsuperscript{th} row and the ($N_{row}/4-1$)\textsuperscript{th} row belong to the subarrays of the bottom and top edges, respectively, we speculate that these two subarrays work together as a single subarray.
%
%
%
We observed that most of DDR4 chips manufactured in $2018$--$2021$ have edge subarrays at every 16K rows ($\times$4: $N_{row}/8$, $\times$8: $N_{row}/4$) or 32K rows ($\times$4: $N_{row}/4$, $\times$8: $N_{row}/2$) boundary regardless of manufacturer.
However, \micron's DDR4 $\times$8 chips manufactured in $2016$ have edge subarrays at every 4K rows ($\times$8: $N_{row}/16$) boundary.
%
The \samsung's HBM2 chip was 32K rows ($N_{row}/4$) and 8K rows ($N_{row}/2$).

Two subarrays working in tandem is reasonable considering the fact that only half of the cells are connected to SAs on either side of the edge (see Figure~\ref{fig:subarray_structure}).
It is aligned with what the prior open bitline structures~\cite{jssc-2001-dummy, isscc-2012-dummy, patent-2015-dummybl} proposed.
Two edge subarrays only connect half the bitlines to the SAs, while the other half bitlines are left as dummies.
Thus, when accessing an edge subarray, a simultaneous access to two rows (one for a pair of edge subarrays) is necessary to form a full single subarray.
%
%

\vspace{-0.04in}
\begin{tcolorbox}[boxsep=0pt,left=3pt,right=3pt,arc=0pt]
\emph{\textbf{Observation-5:}}
For certain DRAM chips with the open bitline structure, two edge subarrays work in tandem to create a single full subarray.
\end{tcolorbox}
\vspace{-0.05in}

\begin{figure}[!tb]
  \center
  \includegraphics[width=0.85\columnwidth]{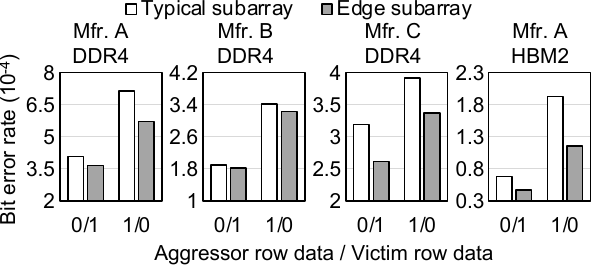}
  \vspace{-0.12in}
  \caption{
    Comparison of \aib-induced BERs based on the subarray type (typical vs. edge subarrays) when the data pattern of an aggressor row is the inverse of a victim row.
  }
  \vspace{-0.1in}
  \label{fig:subarray_error}
\end{figure}

Lastly, we also observed that the edge subarrays demonstrate a uniquely lower bit error rate (BER) for \aib (see Figure~\ref{fig:subarray_error}).
%
%
We tested two different data patterns and measured the \aib-induced BERs of all the subarrays: for (aggressor, victim), (0,1) and (1,0).
%
We identified that for both DDR4 and HBM2, edge subarrays exhibited lower BER, especially when the aggressor data was 1.
We attribute such a difference to the dummy BL of the edge subarrays.
Because only half the BLs are used in the edge subarray, the unused dummy BLs may preserve the precharge voltage state~\cite{isscc-2012-dummy,patent-2015-dummybl}.
This can be backed up by the fact that the BER is lower when the aggressor value is 1, which indicates that the dummy BLs are at least not full $V_{dd}$.

\vspace{-0.04in}
\begin{tcolorbox}[boxsep=0pt,left=3pt,right=3pt,arc=0pt]
\emph{\textbf{Observation-6:}}
Edge subarrays exhibit lower BER from \aib, which can be attributed to dummy BLs.
\end{tcolorbox}
\vspace{-0.05in}

\begin{table}
    \centering
    \small
    \caption{Symbols and Terminologies used throughout \S\ref{sec:5_contribution2}}
    \vspace{-0.05in}
    \label{tbl:5_contribution2}
    \begin{tabular}{p{0.23\columnwidth} p{0.66\columnwidth}} 
        \Xhline{3\arrayrulewidth}
        \textbf{Symbol} & \textbf{Description} \\ 
        \Xhline{1.5\arrayrulewidth}
        \{$i,j$\}       & DRAM cell with WL $i$ and BL $j$. \\
        Top/bottom cell & Type of cells that are isomorphic to each other, as shown in Figure~\ref{fig:cell_notation}.  \\
        \emph{Even/odd WL}     & WL that is indexed by an even/odd number. \\
        $\text{Vic}_\text{0}$    & Tested victim cell for data pattern experiment (\eg, \{1,2\}). \\
        $\text{Vic}_{\text{-}2,\text{-}1,1,2}$    & Adjacent victim cells with distance -2, -1, 1, and 2 (\eg, \{1,0\}, \{1,1\}, \{1,3\}, and \{1,4\}). \\
        $\text{Aggr}_{\text{-}2,\text{-}1,0,1,2}$    & Adjacent aggressor cells with distance -2, -1, 0, 1, and 2 (\eg, \{2,0\}, \{2,1\}, \{2,2\} \{2,3\}, and \{2,4\} for upper aggressor row). \\
        \Xhline{3\arrayrulewidth}
    \end{tabular}
\end{table}

\section{Microscopic Analysis of AIB Characteristics}
\label{sec:5_contribution2}

In this section, we extend our analysis to \emph{microscopic} \aib error characteristics, leveraging the \sixf cell structure and the accurate data swizzling (\textbf{O1}).
Table~\ref{tbl:5_contribution2} summarizes the frequently used symbols and terms throughout this section.
%
%
%

\begin{figure}[!tb]
  \center
  \vspace{0.0in}
  \includegraphics[width=0.85\columnwidth]{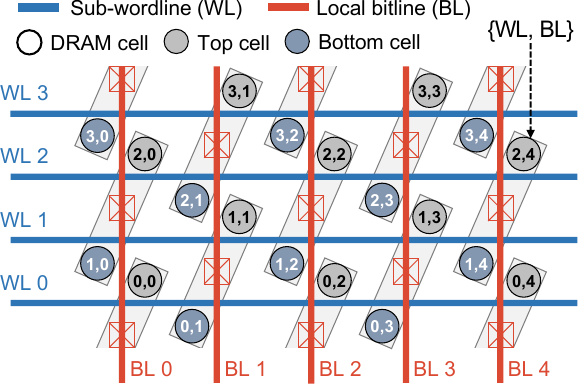}
  \vspace{-0.05in}
  \caption{ 
    \sixf DRAM cell layout of 4x4 cell array. Each cell represents its position through indices of \{WL, BL\}.
  }
  \vspace{-0.05in}
  \label{fig:cell_notation}
\end{figure}

\begin{figure*}[!tb]
  \center
  \vspace{0in}
  \includegraphics[width=0.78\textwidth]{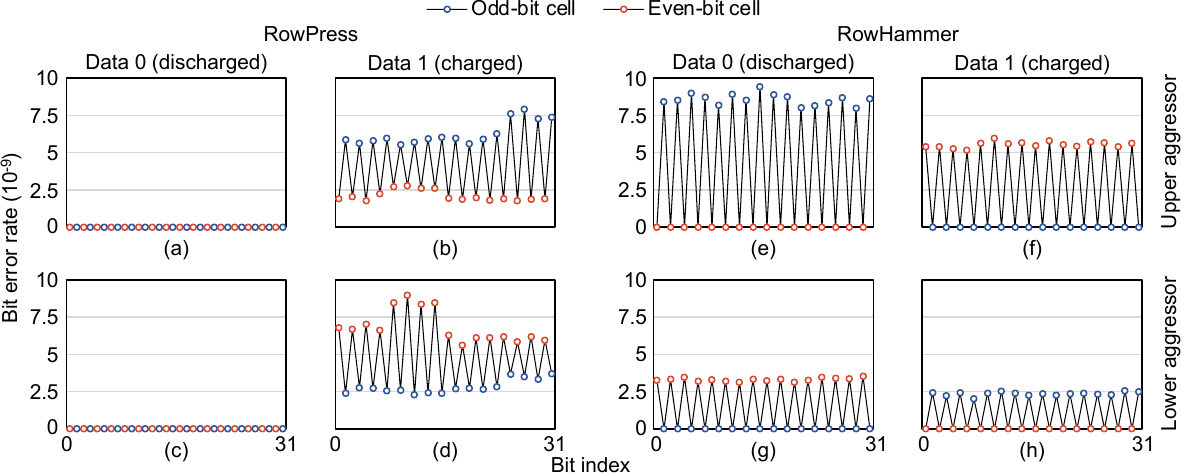}
  \vspace{-0.1in}
  \caption{ 
    Average bit error rate (BER) resulting from (a)-(d) RowPress and (e)-(h) RowHammer in a \samsung DDR4 $\times$4 chip. The errors occur in (b)(d)(f)(h) a charged or (a)(c)(e)(g) a discharged state, induced by (a)(b)(e)(f) an upper or (c)(d)(g)(h) a lower aggressor row. We employ the physically remapped bit index and aggregate the error across repeated 32-bit sequences, taking into account the size of \rdgranul.
  }
  \vspace{-0.12in}
  \label{fig:6f2_graph}
\end{figure*}

\subsection{Top and Bottom Cells in \sixf Structure}
\label{sec:5-1_topbottom}

We adopt a coordinate notation for each cell, to be more concise in our discussion.
For example, \{1,0\} denotes the cell that is connected to $WL1$ and $BL0$ (Figure~\ref{fig:cell_notation}).
Exploiting the regular pattern of cell arrays with a \sixf structure, we categorize all the DRAM cells into two types; top and bottom cells with respect to their relative locations within a P-substrate shared by a pair of cells.
Every top (or bottom) cell is isomorphic to each other.
For a top cell, its upper aggressor row forms a passing gate, whereas its lower aggressor row becomes a neighboring gate.
The opposite holds for a bottom cell.
For instance, the passing gate of the top cell \{1,1\} is the upper $WL2$, whereas its neighboring gate is the lower $WL0$.
Another noticeable pattern is that for a fixed row, the top and bottom cells appear in an alternating manner as the BL index increases (from left to right).
We also define an \emph{even/odd WL}, which refers to WL that is indexed by an even/odd number (\eg, $WL0$ in Figure~\ref{fig:cell_notation} is an \emph{even WL}).

\subsection{\sixf-induced \aib Characteristics}
\label{sec:5-2_aib_characteristics}

For RowPress, we use an attack pattern of 8K single-sided attack with 7.8us for each activation.
For RowHammer, we use 300K single-sided attack\footnote{While double-sided RowHammer attacks can induce more errors with the same number of activation count, it only complicates the error characterization. We used a single-sided attack with enough activation count to compensate for the sufficient number of errors.} with 35ns for each activation.
Based on our experiment, the gradient for flipped cells overlapping with RowPress and RowHammer converges to 0.
%
For the tested 1024 rows, we identified a repeating trend in error with 32-bit granularity.
Thus, Figure~\ref{fig:6f2_graph} reports the average BER when the bit index is modulo 32.
Also, because a victim row with \emph{odd WL} demonstrates a similar yet reversed error pattern with the \emph{even WL} case, we only report the \emph{even WL} case.
%
While the figure only summarizes the tested results of \samsung-2021 DDR4, similar behavior was observed in other manufacturers and HBM2.
%
%
%
We emphasize that our analysis is highly dependent on accurate data swizzling reverse-engineering.

\noindent
\textbf{RowPress:}
Our RowPress experiment enabled us to successfully observe an alternating error pattern as expected.
Figure~\ref{fig:6f2_graph}(a,b,c,d) reports the BER rate of a certain fixed (\emph{even WL}) victim row, for the upper or lower aggressor row.
%
%
RowPress-induced bitflip was observed only in victim data 1 (charged state), as reported before~\cite{isca-2023-rowpress}.
In the charged state, we observed an alternating BER pattern as the bit index (BL index) increased.
This accurately reflects the fact that top and bottom cells appear to take turns within a victim row (Figure~\ref{fig:cell_notation}), meaning that passing and neighboring gates appear in an alternating way, for a fixed aggressor direction (\eg, upper).
Moreover, the BER pattern is \emph{reversed} when the direction of the aggressor changes (upper and lower) or the victim row changes (\emph{even} and \emph{odd WL}).
Such inversion can be explained by the reversed-symmetrical structure of \sixf.
%

\vspace{-0.04in}
\begin{tcolorbox}[boxsep=0pt,left=3pt,right=3pt,arc=0pt]
\emph{\textbf{Observation-7:}}
RowPress occurs in an alternating pattern, which is in a reversed form between the upper/lower aggressor and \emph{even/odd WL} victim row.
\end{tcolorbox}
\vspace{-0.05in}

\noindent
\textbf{RowHammer:}
We also recognized an alternating error pattern in the RowHammer experiment.
Figure~\ref{fig:6f2_graph}(e,f,g,h) reports the BER rate of a particular (\emph{even WL}) victim row for the upper and lower aggressor row.
Examining the charged state victim row, we again observe a similar trend of alternating BER as the bit index increases.
Also, such alternation is reversed between the upper/lower aggressor row and the \emph{even/odd WL} victim row.
Similar alternation is observed for the discharged state as well, yet in a reversed form.

\vspace{-0.04in}
\begin{tcolorbox}[boxsep=0pt,left=3pt,right=3pt,arc=0pt]
\emph{\textbf{Observation-8:}}
RowHammer occurs in an alternating pattern, in a reversed form between the upper/lower aggressor row, \emph{even/odd WL} victim row, and charged/discharged victim row.
\end{tcolorbox}
\vspace{-0.05in}

\begin{figure}[!tb]
  \center
  \vspace{0.0in}
  \includegraphics[width=0.75\columnwidth]{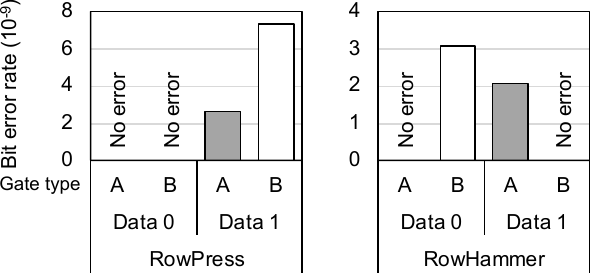}
  \vspace{-0.1in}
  \caption{
  BER resulting from RowPress and RowHammer. The errors occur in a charged state or a discharged state, induced by a gate type A or B. A/B gates can be either passing/neighboring or neighboring/passing gates.
  }
  \vspace{-0.1in}
  \label{fig:6f2_graph_summary}
\end{figure}

Examining the gate types for each of the four tested situations (data 1/0 and upper/lower aggressor), we recognized that RowHammer can also happen on two gate types (see Figure~\ref{fig:6f2_graph_summary}).
We denote the gate types as A and B because we could not fully determine whether A is a passing and B is a neighboring gate or the opposite, unless we refer to the prior study.\footnote{
Prior works~\cite{iedm-2017-reliability, ted-2020-passing-gate} claimed that the failure mechanisms of RowHammer are charge injection from either neighboring gate (NG) or passing gate (PG), when victim cell's data is 1 or 0, respectively.
However, considering the relationship between charge state of the victim cell and gate type, the characteristics of RowPress are opposite to those of RowHammer, which is different from the previous study~\cite{arxiv-2023-dsac}.
Therefore, it is difficult to determine the gate type.}
%
Moreover, we discovered that, against RowHammer, each cell is only affected by one type of gate type at a time, which is reversed when written data changes.
For example, the cell with a bit index 0 demonstrates susceptibility against the upper aggressor row for data 1, and against the lower aggressor row when data is 0.
Considering that the gate types on the upper and lower are the opposite for both the top and bottom cells, we can conclude that (1) RowHammer occurs in both gate types and (2) the susceptible gate type is reversed when written data changes.

\vspace{-0.04in}
\begin{tcolorbox}[boxsep=0pt,left=3pt,right=3pt,arc=0pt]
\emph{\textbf{Observation-9:}}
RowHammer occurs at both the neighboring and passing gates.
\end{tcolorbox}
\vspace{-0.05in}

\vspace{-0.04in}
\begin{tcolorbox}[boxsep=0pt,left=3pt,right=3pt,arc=0pt]
\emph{\textbf{Observation-10:}}
Against RowHammer, a victim cell is only susceptible to one gate type (passing/neighboring) at a time, which is reversed when written data changes (charged/discharged).
\end{tcolorbox}
\vspace{-0.05in}

\subsection{\aib Data Pattern Dependence}
\label{sec:5-3_column_data_pattern}

Based on our reverse-engineering of data swizzling and \sixf, we execute a detailed study on the data pattern dependence on \aib.
Also, we propose an adversarial data pattern of victim/aggressor rows given a single victim cell based on the analysis.
While prior studies~\cite{isca-2014-flipping, isca-2020-revisit, sp-2020-rambleed, asiaccs-2019-pinpoint, micro-2021-deeper} considered several types of data patterns, they were not based upon an accurate data swizzling. 
This resulted in inaccurate data pattern mapping (\S\ref{sec:3-3_common_pitfalls}), resulting in misinterpretations.

\begin{figure}[!tb]
  \center
  \vspace{0in}
  \includegraphics[width=0.95\linewidth]{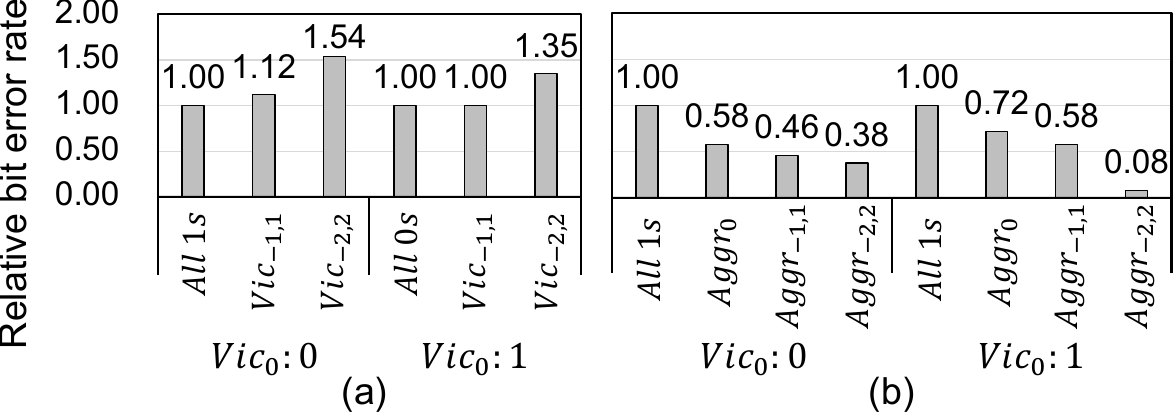}
  \vspace{-0.05in}
  \caption{
    Relative BER influenced by (a) victim cells at distance-one and distance-two and (b) aggressor cells at distance-zero, distance-one, and distance-two, ordered in a descending manner.
  }
  \vspace{-0.12in}
  \label{fig:cell_influence}
\end{figure}

\noindent
\textbf{Victim row horizontal data pattern:}
We observed that for a particular victim cell of $\text{Vic}_0$ (\eg, \{1,2\} in Figure~\ref{fig:cell_notation}), its adjacent four victim cells of $\text{Vic}_{\text{-}2,\text{-}1,1,2}$ (\eg, \{1,0\}, \{1,1\}, \{1,3\}, and \{1,4\}, respectively) affect the RowHammer vulnerability.
%
%
%
Figure~\ref{fig:cell_influence}(a) summarizes the result, whose baseline BER is when all cells of aggressor and victim rows are in either (zeros, ones) or (ones, zeros).
When testing the impact of four adjacent victim cells, we did not alter the data of the aggressor row.
First, we recognized that BER is the worst when all four adjacent victim cells
%
$\text{Vic}_{\text{-}2,\text{-}1,1,2}$ hold the opposite value of the $\text{Vic}_0$.
Also, the impact of distance-two cells 
$\text{Vic}_{\text{-}2,2}$
is more significant than the impact of distance-one adjacent victim cells 
$\text{Vic}_{\text{-}1,1}$.
%
For example, when cell 
$\text{Vic}_0$ 
was 0, altering the value of cells 
$\text{Vic}_{\text{-}1,1}$ 
resulted in 1.12$\times$ BER, whereas changing $\text{Vic}_{\text{-}2,2}$ caused 1.54$\times$ BER.
When $\text{Vic}_0$ was 1, the increase in BER was 1.00$\times$ and 1.35$\times$, respectively.
We explain this to be attributed to the \sixf structure, where the \emph{physical} distance difference between $\text{Vic}_{\text{-}1,1}$ and $\text{Vic}_{\text{-}2,2}$ is smaller than 2$\times$, and physical distance between $\text{Vic}_{\text{-}1,1}$ is also not identical (see Figure~\ref{fig:cell_notation}).

\vspace{-0.04in}
\begin{tcolorbox}[boxsep=0pt,left=3pt,right=3pt,arc=0pt]
\emph{\textbf{Observation-11:}}
Given a particular victim cell ($\text{Vic}_{0}$), its horizontally adjacent four victim cells ($\text{Vic}_{\text{-}2,\text{-}1,1,2}$) affect its RowHammer vulnerability, which is the strongest in $\text{Vic}_{\text{-}2,2}$.
\end{tcolorbox}
\vspace{-0.05in}

\noindent
\textbf{Aggressor row horizontal data pattern:}
We observed a horizontal data pattern dependence on RowHammer on the aggressor row.
Given a particular victim cell ($\text{Vic}_{0}$), we denote the directly adjacent aggressor cell as $\text{Aggr}_0$ (\eg, \{2,2\}) and its four adjacent aggressor cells as $\text{Aggr}_{\text{-}2,\text{-}1,1,2}$ (\eg, \{2,0\}, \{2,1\}, \{2,3\}, and \{2,4\}).
While previously recognized data pattern dependence was mainly limited to $\text{Aggr}_0$, we newly discovered that $\text{Aggr}_{\text{-}2,\text{-}1,1,2}$ impact the RowHammer bitflip.
In fact, $\text{Aggr}_{\text{-}2,\text{-}1,1,2}$ are more influential than $\text{Vic}_{\text{-}2,\text{-}1,1,2}$.
With the baseline of aggressor and victim row of (zeros, ones) and (ones, zeros), we measured the BER of $\text{Vic}_0$ while only varying the value of $\text{Aggr}_{\text{-}2,\text{-}1,1,2}$ (Figure~\ref{fig:cell_influence}(b)).
Changing $\text{Aggr}_0$ value decreased BER by 0.58$\times$ (0.72$\times$) when $\text{Vic}_0$ was 0 (1).
Altering $\text{Aggr}_{\text{-}1,1}$ and $\text{Aggr}_{\text{-}2,2}$ resulted in 0.46$\times$ (0.58$\times$) and 0.38$\times$ (0.08$\times$) drop in BER for $\text{Vic}_0$ of 0 (1).
Unlike the victim cell cases, the influence of horizontally adjacent aggressor cells was the largest when it was closest to $\text{Vic}_0$.

\vspace{-0.04in}
\begin{tcolorbox}[boxsep=0pt,left=3pt,right=3pt,arc=0pt]
\emph{\textbf{Observation-12:}}
Given a particular victim cell ($\text{Vic}_0$), not only its closest aggressor cell ($\text{Aggr}_0$) but also horizontally adjacent aggressors ($\text{Aggr}_{\text{-}2,\text{-}1,1,2}$) impact the RowHammer bitflip, which is the strongest when physically closest.
\end{tcolorbox}
\vspace{-0.05in}

\begin{figure}[!tb]
  \center
  \vspace{-0.0in}
  \includegraphics[width=0.8\columnwidth]{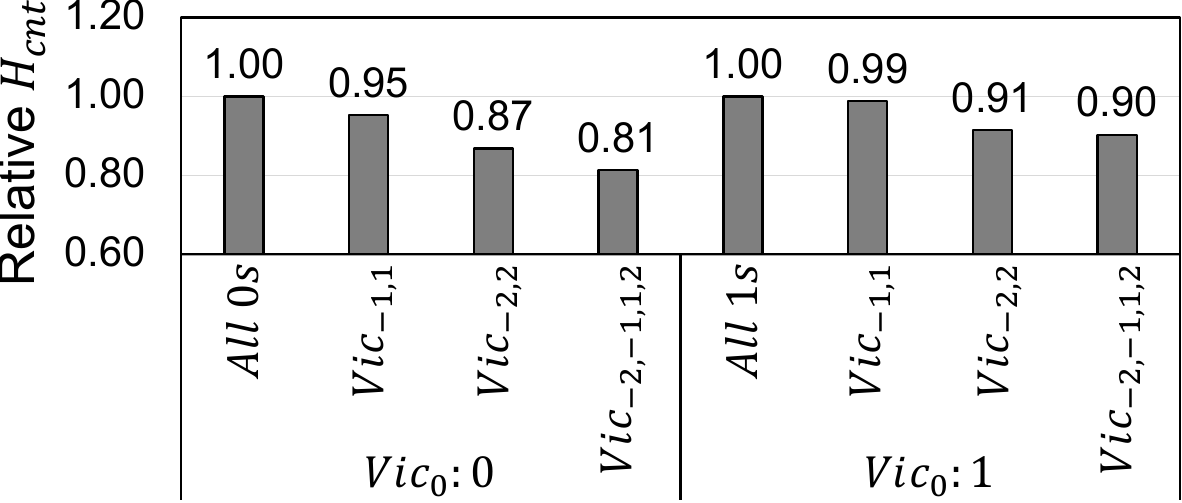}
  \vspace{-0.1in}
  \caption{
    Relative $\text{H}_{\text{cnt}}$ as data changes in the other victim cells.
    $\text{Vic}_n$ means that data different from $\text{Vic}_0$ was written to a cell n away from $\text{Vic}_0$.
    The aggressor row is all 1s (0s) when $\text{Vic}_0$ = 0 (1).
  }
  \vspace{-0.1in}
  \label{fig:victimcell_experiment}
\end{figure}

\begin{figure}[!tb]
  \center
  \vspace{0in}
  \includegraphics[width=1\linewidth]{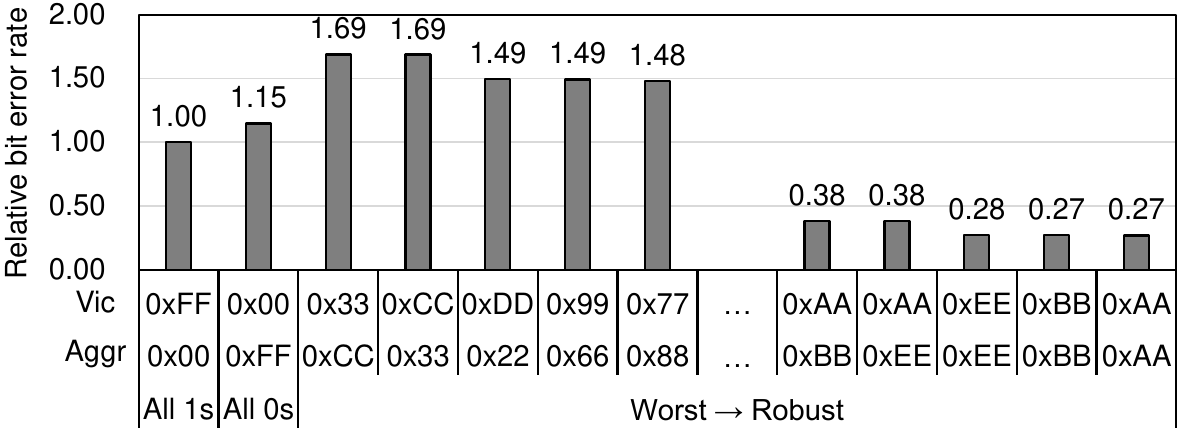}
  \vspace{-0.05in}
  \caption{
    BER according to the data pattern of the aggressor and victim rows. Both the aggressor and victim row exhibit a repeating 4-bit sequence pattern. We represent the data pattern with values actually written to the MAT.
  }
  \vspace{-0.12in}
  \label{fig:pattern_experiment}
\end{figure}

\subsection{Adversarial Data Pattern for \hcnt and BER}
\label{sec:5-4_worst-case data pattern}

Leveraging the accurate data swizzling and horizontal data pattern dependence, we introduce an adversarial data pattern in terms of minimum activation count to cause the first bitflip (\hcnt) and BER of the whole victim row.

\noindent\textbf{Adversarial data pattern for \hcnt:}
We identified that setting $\text{Vic}_\text{-2,-1,1,2}$ and $\text{Agg}_\text{-2,-1,0,1,2}$ as the opposite value of $\text{Vic}_\text{0}$ deteriorates the \hcnt value by up to 0.81$\times$.
Due to the definition of \hcnt, the adversary can only target a particular victim cell, $\text{Vic}_0$, instead of the whole victim row.
Compared to the baseline \hcnt where victim and aggressor row values are both 0s (1s), setting the $\text{Vic}_{\text{-}1,1}$, $\text{Vic}_{\text{-}2,2}$, and $\text{Vic}_{\text{-}2,\text{-}1,1,2}$ to the opposite value of 1s (0s) decreased \hcnt by 0.95$\times$ (0.91$\times$), 0.87$\times$ (0.91$\times$), and 0.81$\times$ (0.90$\times$), respectively (see Figure~\ref{fig:victimcell_experiment}).
The fact that \hcnt decreases by a larger amount for $\text{Vic}_{\text{-}2,2}$ than $\text{Vic}_{\text{-}1,1}$ concurs with prior observation \textbf{(O11)}.
%

\vspace{-0.1in}
\begin{tcolorbox}[boxsep=0pt,left=3pt,right=3pt,arc=0pt]
\emph{\textbf{Observation-13:}}
    \hcnt is lowered by up to 0.81$\times$, when all cells of $\text{Vic}_{\text{-}2,\text{-}1,1,2}$ and $\text{Aggr}_{\text{-}2,\text{-}1,0,1,2}$ hold the opposite value of $\text{Vic}_0$.
\end{tcolorbox}
\vspace{-0.05in}

%
%

\begin{figure}[!tb]
  \center
  \vspace{0in}
  \includegraphics[width=0.83\columnwidth]{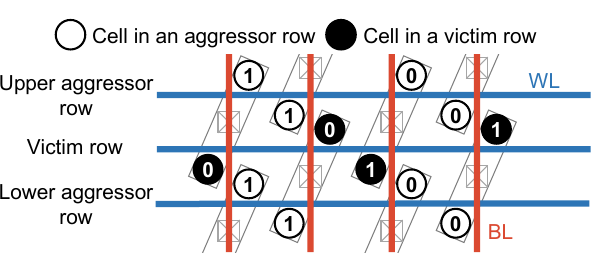}
  \vspace{-0.09in}
  \caption{
    Among the data patterns where the data in the victim rows and the aggressor rows repeat 4-bit each, the worst-case data pattern can cause the largest number of RowHammer-induced bitflips in the victim row.
    The data of the aggressor row and the victim row are opposite, each data repeating two 0s or 1s.
    The data of the same bit index in the victim row and the aggressor rows are reversed, and two 0s and two 1s are repeated in each row.
  }
  \vspace{-0.09in}
  \label{fig:worst-case}
\end{figure}

\noindent\textbf{Adversarial data pattern for BER:}
We identified an adversarial aggressor and victim row data pattern that deteriorates the overall BER of the victim row by up to 1.69$\times$.
We sweep 16 different data arrangements that have repeated 4-bit patterns for both victim and aggressor, testing 256 combinations in total.
Figure~\ref{fig:pattern_experiment} summarizes the result.
The baseline is the BER when the victim and aggressor row have $\mathtt{0xFF}$ and $\mathtt{0x00}$ patterns, respectively.
%
Among the tested 256 combinations, the worst case was when the victim and aggressor rows have $\mathtt{0x33}$ and $\mathtt{0xCC}$, resulting in 1.69$\times$ higher BER than the baseline (Figure~\ref{fig:worst-case}).
Noticeably, this adversarial pattern is when the vertically adjacent aggressor and victim cells (\eg, $\text{Vic}_\text{i}$ and $\text{Aggr}_\text{i}$) hold the opposite value, with a repeating pattern of 2-bits.
The reason why the two-bit repeating pattern is worse than the one-bit alternating pattern can be explained by \textbf{(O11)}, or that $\text{Vic}_{\text{-}2,2}$ is more influential than $\text{Vic}_{\text{-}1,1}$.

\vspace{-0.04in}
\begin{tcolorbox}[boxsep=0pt,left=3pt,right=3pt,arc=0pt]
\emph{\textbf{Observation-14:}}
    The overall BER from \aib-induced bitflips can deteriorate by up to 1.69$\times$ when the victim and aggressor rows hold data pattern that repeats $\mathtt{0x33}$ and $\mathtt{0xCC}$. 
\end{tcolorbox}
\vspace{-0.05in}

\section{New Vulnerabilities and Protections}
\label{sec:6_vuln_prot}

In this section, we first examine the impact of our findings on existing \aib attacks and defenses.
%
We propose a simple yet effective \aib protection solution against our findings.
%
We also investigate the security implications and use cases of our observations at both macroscopic and microscopic levels.

\subsection{Exacerbating Known \aib Attacks}
\label{sec:6-1_worsened_aib}

\noindent
\textbf{Coupled-row activation:} 
Coupled-row activation \textbf{(O3)} can circumvent existing \aib defense mechanisms by breaking their assumption that bitflips only occur in the rows adjacent to the tracked aggressor rows.
%
%
We can envision the following scenario.
Suppose the two coupled rows are denoted as row-A and row-B.
When an attacker only activates row-A, the \aib protection scheme is likely to only track row-A, especially when the solution is located at an MC.
If the protection solution is victim row refresh-based~\cite{micro-2020-graphene}, it can still be secure by unintentionally refreshing victims of row-B.
However, when the protection solution is based on state-of-the-art MC-side row swapping~\cite{asplos-2022-rrs,hpca-2023-srs}, it can be neutralized because it will only relocate row-A, without affecting row-B.

Coupled-row activation also deceives the existing activation counter structures.
When an attacker splits DRAM activations into each of a coupled-row pair respectively (\eg, row-A and row-B being coupled), the counter structures perceive attacker's activations as two different row activations.
However, two rows in a coupled-row pair compose a single DRAM row; thus, an attacker can easily bypass the counter structure by splitting the activations into the two rows in a coupled-row pair.
%
Even if the MC-side \aib protection scheme acquires the coupled-row information, although itself challenging~\cite{irps-2022-secrecy}, the area or performance cost might become prohibitive.
Most \aib mitigation schemes adopt SRAM- or CAM-based row tracking structures~\cite{isca-2014-flipping,isca-2019-twice,micro-2020-graphene,hpca-2022-mithril}.
Coupled-row activation doubles the number of rows to track (performance cost from twice the mitigative actions) or effectively doubles the number of activations (area cost from a larger tracking table).

Coupled-row activation aids the \aib attackers from the system perspective.
Most \aib attacks utilize memory templating/massaging techniques~\cite{sec-2016-fengshui,ccs-2016-drammer,sp-2020-rambleed,sec-2021-smash} to prepare the \aib (hammering) phase.
In the memory massaging/templating phase, an attacker controls the system memory space to locate the target (victim) memory page to conduct \aib (hammering).
Coupled-row activation increases the probability of successful memory massaging as the attacker concurrently accesses a single attacker-controlled memory page and benign pages belonging to other processes.
Accordingly, coupled-row activation ensures a higher probability of guaranteeing adjacency between the attacker and victim pages, posing a severe threat.
%

\noindent
\textbf{\aib adversarial pattern:} 
The \aib adversarial data pattern we propose worsens the existing \aib attacks 1) by decreasing the effective activation counts for an attack and 2) providing complex data pattern dependence of \aibs.
Also, existing data pattern-aware \aib attacks~\cite{sp-2020-rambleed,asiaccs-2019-pinpoint} should be modified to successful attacks.
Pinpoint RowHammer~\cite{asiaccs-2019-pinpoint} and RAMBleed~\cite{sp-2020-rambleed} assume \aibs are only affected by row-wise (vertical) data patterns.
However, our findings suggest that the influence of the column-wise (horizontal) data pattern should be considered.
Throughout this, it is possible to increase the accuracy of the existing data pattern-aware \aib attacks.

\subsection{Revising Existing \aib Protection}
\label{sec:6-2_revising_existing}

\noindent
\textbf{Protecting coupled-row activation:}
To prevent bitflips caused by coupled-row activation, existing \aib defense mechanisms must consider a coupled row for each activation.
Based on Table~\ref{tbl:4_subarray_size}, the relationship between coupled rows can be expressed using a simple calculation (\eg, $(n, n+64K)$ for $n:\{n \in \mathbb{N}| 0 \le n < 64K \}$).
Therefore, the existing tracking-based mitigation should consider coupled rows by additionally tracking its coupled row for each activation.
Furthermore, considering the varying relationships with coupled rows per DRAM, in-DRAM \aib mitigation will be promising to handle the coupled-row activation.

The recently introduced Directed Refresh Management (DRFM) command is one of the prominent solutions for protecting coupled-row activation.
The DDR5 JEDEC standard introduced DRFM commands as an \aib mitigation~\cite{jedec-ddr5}.
DRFM works as follows: An MC samples the activated DRAM row on \texttt{PRE} commands, and DRAM stores the corresponding DRAM row address.
Then, when a DRAM receives the \texttt{DRFM} command, DRAM refreshes the physically adjacent rows with the stored row address.
There have been attempts to utilize an RFM interface to mitigate \aib attacks~\cite{hpca-2022-mithril, sp-2022-protrr, hpca-2023-shadow}.
Existing RFM-based \aib mitigation methods track DRAM activation commands and efficiently send an RFM command to trigger in-DRAM \aib mitigation actions.
If the DRAM manufacturers disclose the coupled-row relationship information in either the DRAM chip's mode register or the DRAM module's Serial Presence Detect chip, an MC can read the information using Mode Register Read commands or \texttt{i2c} interface.
Then, an MC can effectively track both coupled-row activations as a single aggressor row's activation.
Thus, coupled-row-related \aib attacks can be mitigated by adopting a DRFM-based \aib mitigation with a minimum modification.

\noindent
\textbf{Protecting adversarial data pattern \aib attacks:} 
Data scrambling~\cite{patent-2011-scrambling,hpca-2017-coldboot,amd-encryption} can be an efficient mitigation technique to defend against adversarial data pattern \aib attacks.
Data scrambling obfuscates the data written to the memory devices by masking the original data with the randomly generated bitmask or using encryption algorithms.
Originally, Intel proposed data memory scrambling to enhance the resistance to irregular signals and power noise.
Recently, it has been revealed that data scrambling can be used for security enhancements~\cite{eprint-2016-sgx,hpca-2017-coldboot,amd-encryption}.
Modern processors manufactured by two major processor vendors (\eg, Intel and AMD) enable the MC-side memory scrambling or encryption by default~\cite{patent-2011-scrambling,hpca-2017-coldboot,amd-encryption}.
The adversarial data pattern we propose consists of both row-wise and column-wise data pattern \textbf{(O11-14)}.
Therefore, through a more robust PRNG algorithm involving both row and column addresses for generating bitmasks, the memory scrambling method can guarantee strong \aib adversarial data pattern resistance as in the case of the cold boot attacks~\cite{hpca-2017-coldboot}.
Lastly, adversarial data pattern-aware error correcting code (ECC) algorithm/design and coding theory could be promising mitigation techniques~\cite{memsys-2017-blcrosstalk,hpca-2018-synergy,hpca-2022-safeguard,micro-2023-cube,memsys-2023-rampart}.


\subsection{Possible New Attacks and Use Cases of Our Findings}
\label{sec:6-3_new_threat}

Our findings on DRAM microarchitectures and operations introduce new threats to DRAM-based memory systems.
The presence of edge subarrays and coupled-row activation leads to differences in power consumption based on the DRAM and subarray type.
The activation of edge subarrays triggers two activations of rows in each different subarray, doubling the DRAM dynamic power consumption.
Similarly, coupled-row activation doubles the power consumption due to the activation occurring in an arbitrary row.
If DRAM power consumption is analyzed, it is possible to distinguish which memory (row and subarray) is accessed.
Thus, an additional analysis of DRAM power-based side-/covert-channel could be intriguing~\cite{ccs-2022-hammerscope}.

Also, our findings enhance the reliability and robustness of processing in memory (PIM) techniques.
For example, well-known PIM techniques, such as in-memory row copy operations (RowCopy)~\cite{micro-2013-rowclone,micro-2019-computedram} or bitwise in-memory operation using many row activation~\cite{micro-2019-computedram,micro-2022-fracdram}, pose a significant threat to computer systems.
For example, considering coupled-row activation, RowCopy operations in some $\times$4 DRAM chips or HBM2 cause unauthorized data copy.
Unauthorized data copy of unintended DRAM rows compromises the confidentiality of modern computer systems.
Therefore, processing using memory~\cite{micro-2013-rowclone} necessitates a deeper understanding of precise DRAM operations.
We believe that future DRAM-based memory system research must consider DRAM microarchitectures, associated operations, and characteristics for security.
\section{Related Work}
\label{sec:7_related_work}

There has been a large body of DRAM experimental analysis and characterization research, such as analyzing retention time variation~\cite{isca-2013-retention} and latency variation~\cite{pomacs-2017-dramlatency} in commercial DRAM chips.
After the advent of DRAM read disturbance error (RowHammer), a number of works have attempted to analyze and characterize the DRAM \aib characteristics~\cite{sp-2020-suscep,isca-2020-revisit,micro-2021-uncovering}.
Cojocar et al.~\cite{sp-2020-suscep} provide the method to evaluate the \aib vulnerability in the cloud environments.
Kim et al.~\cite{isca-2020-revisit} demonstrate an experimental characterization of \aib on real DRAM chips and evaluate \aib mitigation techniques based on their characterization results.
Hassan et al.~\cite{micro-2021-uncovering} uncover undocumented in-DRAM TRR mechanisms on real DRAM chips.
Also, there are works that experimentally analyze the effects of various factors (such as temperature~\cite{micro-2021-deeper, ccs-2022-hammerscope, arxiv-2023-spyhammer} and wordline voltage~\cite{dsn-2022-wlvoltage}) on \aib characteristics.
A new type of \aib attack called RowPress~\cite{isca-2023-rowpress} was recently introduced.
However, to the best of our knowledge, \name, which extended~\cite{cal-2023-xray}, is the first work that deeply considers DRAM \sixf cell structure and exact DRAM internal mappings on DRAM \aib characterization.

\section{Conclusion}
\label{sec:8_conclusion}

We have reliably revealed the DRAM microarchitectures, associated behaviors, and activate-induced bitflip (\aibs) characteristics through \aib tests, retention time tests, and RowCopy using commercial DRAM chips.
We showed that precise mapping information of DRAM modules and chips is necessary to accurately analyze the \aib characteristics.
We discovered undisclosed DRAM microarchitectures and associated behaviors, such as dummy bitline, edge subarray, and coupled-row activation.
We clarified the common misconceptions from prior DRAM studies, such as the non-adjacent \aib phenomenon and fixed height of subarrays.
By considering the DRAM's microscopic aspect, such as DRAM \sixf cell structure, we also identified the data pattern dependency on the \aib phenomenon.
We anticipate that our new observations, clarifications, and the experimental methodology will enrich future DRAM \aib experimental analyses, and \aib attacks and defenses.

\section*{Acknowledgements}
This work was supported by Samsung Electronics Co., Ltd (IO201207-07812-01),  an Institute of Information \& communications Technology Planning \& Evaluation (IITP) grant funded by the Korea government (MSIT) (No. 2020-0-01300, No. 2021-0-01343, and IITP-2023-RS-2023-00256081), and a grant from PRISM, one of the seven centers in JUMP 2.0, a Semiconductor Research Corporation (SRC) program sponsored by DARPA.
Nam Sung Kim has a financial interest in Samsung Electronics. 
The EDA tool was supported by the IC Design Education Center (IDEC), Korea.
The ICT at Seoul National University (SNU) provides research facilities.
Jung Ho Ahn, the corresponding author, is with the Department of Intelligence and Information and the Interdisciplinary Program in Artificial Intelligence, SNU.

%
%
%
%
%


\appendix
\section*{Artifact appendix}
\label{A_artifact_appendix}
\subsection{Abstract}

Our artifact provides guidelines, source code, and scripts for reproducing the figures in the paper.
We offer FPGA-based infrastructures for experiments, including modified SoftMC~\cite{hpca-2017-softmc} and DRAM Bender~\cite{tcad-2023-dram-bender}.
%
The experiments of \name consist of 1) RowHammer attack, 2) RowCopy operation, 3) Retention time test, and 4) RowPress attack for DDR4 and HBM2.
We provide scripts to analyze the experimental results and plot the figures presented in this paper.

\subsection{Artifact check-list (meta-information)}
\label{}

{\small
\begin{itemize}
  \item {\bf Program:}  
  
  FPGA-based infrastructure: Modified SoftMC~\cite{hpca-2017-softmc} and DRAM Bender~\cite{tcad-2023-dram-bender} for Table~\ref{tbl:5_contribution2} and Figure~\ref{fig:subarray_error}, \ref{fig:6f2_graph}, \ref{fig:6f2_graph_summary}, \ref{fig:pattern_experiment}.
  

  \item {\bf Compilation:} C++14 for FPGA-based infrastructure and Python 3.6.9 for analysis.
  
  \item {\bf Metrics:} Address remapping and bit error rate.
  
  \item {\bf Output:} \textit{CSV} files are generated from FPGA experiments.
  \textit{PNG} and \textit{SVG} files are graphs, similar to the figures in this paper, generated by scripts.
  
  \item {\bf How much disk space required (approximately)?:} 2.2GB for an FPGA-based infrastructure and 1GB for the results of \name.
  
  \item {\bf How much time is needed to prepare workflow (approximately)?:} 1 hour.
  
  \item {\bf How much time is needed to complete experiments (approximately)?:} 5 hours.
  
  \item {\bf Publicly available?:} Yes.
  
  \item {\bf Code licenses (if publicly available)?:} The MIT License for prior works (\ie, SoftMC~\cite{hpca-2017-softmc} and DRAM Bender~\cite{tcad-2023-dram-bender}) and \name.
  
  \item {\bf Archived?:} Yes. \url{https://zenodo.org/records/11044630} 
  
\end{itemize}
}

\subsection{Description}

\subsubsection{How to Access}
Our modified FPGA-based infrastructure, script files, and instructions of our experiments are publicly available at GitHub repository ({\small{\url{https://github.com/scale-snu/AE_DRAMScope_ISCA2024}}}) and Zenodo ({\small{\url{https://zenodo.org/records/11044630}}}).
Prior works for our FPGA-based infrastructure are publicly available at: 

{\small
\begin{itemize}
  \item {\bf SoftMC:} \url{https://github.com/CMU-SAFARI/SoftMC}
  \item {\bf DRAM Bender:} \url{https://github.com/CMU-SAFARI/DRAM-Bender}
\end{itemize}
}

\subsubsection{Hardware Dependencies}
We utilize the following FPGA-based infrastructure:
\begin{itemize}
    \item {A host x86 machine supporting PCIe 3.0 $\times$16 slots}
    \item {An FPGA board with DIMM slots supported by DRAM Bender~\cite{tcad-2023-dram-bender} (e.g., Xilinx Alveo U200~\cite{u200} and U280~\cite{u280})}
    \item {An FPGA board with HBM2 supported by modified SoftMC~\cite{hpca-2017-softmc} for HBM2 (e.g., Xilinx Alveo U280~\cite{u280})}
    \item {A rubber heater attached to the DIMM for temperature control}
    \item {A temperature controller connected to the rubber heater} \\
\end{itemize} 

We conducted experiments using Intel Core i5-7500 (Kaby Lake) CPU and i5-8400 (Coffee Lake) CPU.
We experimented with RDIMMs containing x4 and x8 chips, utilizing RDIMMs from Samsung (e.g., \textit{M393A2K40BB1-CRC}), SK Hynix (e.g., \textit{HMA84GR7JJR4N-WM}), and Micron (e.g., \textit{MTA18ASF2G72PZ-2G9}).

\subsubsection{Software Dependencies}
\begin{itemize}
    \item {GNU Make 4.1+}
    \item {C++14 build toolchain}
    \item {Python 3.6.9+}
    \item {AMD Vivado 2020.2+}
    \item {pip packages: \textit{matplotlib} and \textit{seaborn}}
    \item {Ubuntu 18.04 (Linux kernel 5.4.0-150-generic)}
\end{itemize}

\subsection{Installation and Experiment Workflow}
For more detailed guidelines, please refer to \texttt{README.md} files in the following repository:
{\small \url{https://github.com/scale-snu/AE_DRAMScope_ISCA2024}}

\subsection{Evaluation and Expected Results}
After conducting each experiment on an FPGA-based infrastructure, the results files, including addresses where bitflips occurred or other experiment results, are automatically generated.
We use scripts to generate the figures in this paper based on the data in the result files. 
Please refer to the \texttt{README.md} files in the repository for the detailed processes.

\balance
\bibliographystyle{IEEEtranS}
\bibliography{ref}

\end{document}